\documentclass[10pt,journal,compsoc]{IEEEtran}
\usepackage{tabularx}
\usepackage{tcolorbox}
\usepackage{tabu}
\usepackage{multirow}
\usepackage{graphicx}
\usepackage{makecell}
\usepackage{subfigure}
\usepackage[figuresright]{rotating}
\usepackage{booktabs}
\usepackage{amsmath}
\usepackage{pifont}
\usepackage{longtable}
\usepackage{wasysym}
\usepackage{lscape}
\usepackage{supertabular}
\usepackage{color}
\usepackage{soul}
\usepackage{listings}
\usepackage{xspace}
\usepackage{diagbox}
\usepackage{ragged2e}
\usepackage{enumitem}
\usepackage{threeparttable}
\usepackage{balance}
\usepackage{url}
\usepackage{cite}
\usepackage{textcomp}
\usepackage{xcolor}
\usepackage{wrapfig}
\usepackage{subcaption}
\usepackage{pdflscape}
\usepackage{hyperref}
\usepackage{tikz}
\usepackage[normalem]{ulem}
\usepackage{amssymb}
\usepackage{xcolor,colortbl}

\newcommand{\tabincell}[2]{\begin{tabular}{@{}#1@{}}#2\end{tabular}}

\newcommand{\find}[1]{
\begin{tcolorbox}[leftrule=1mm,toprule=0mm,bottomrule=0mm,left=1pt,right=2pt,top=2pt,bottom=2pt]
\em #1
\end{tcolorbox}
}

\definecolor{lightgray}{gray}{0.8}

\newcommand{\intuition}[1]{
\begin{tcolorbox}[colback=white,boxrule=1pt,top=0pt,bottom=0pt,left=1pt,right=2pt,top=2pt,bottom=2pt]
\em #1
\end{tcolorbox}
}

\newboolean{showcomments}
\setboolean{showcomments}{true}
\ifthenelse{\boolean{showcomments}}
 { \newcommand{\mynote}[2]{
      \fbox{\bfseries\sffamily\scriptsize#1}
        {\small$\blacktriangleright$\textsf{\emph{#2}}$\blacktriangleleft$}}}
        { \newcommand{\mynote}[2]{}}

\newcommand{\xy}[1]{\textcolor{black}{#1}}

\newcolumntype{L}[1]{>{\raggedright\let\newline\\\arraybackslash\hspace{0pt}}m{#1}}
\newcolumntype{C}[1]{>{\centering\let\newline\\\arraybackslash\hspace{0pt}}m{#1}}
\newcolumntype{R}[1]{>{\raggedleft\let\newline\\\arraybackslash\hspace{0pt}}m{#1}}

\usepackage{tikz}
\newcommand*\circled[1]{\tikz[baseline=(char.base)]{
            \node[shape=circle,draw,inner sep=1.5pt] (char) {#1};}}

\usepackage[colorinlistoftodos]{todonotes}

\begin{document}

\title{Multitask-based Evaluation of Open-Source LLM on Software Vulnerability}

\author{Xin Yin,
        Chao Ni$^{\star}$,
        and Shaohua Wang
\thanks{Both Xin Yin and Chao Ni are with the State Key Laboratory of Blockchain and Data Security, Zhejiang University, Hangzhou, China. 
Chao Ni is also with Hangzhou High-Tech Zone (Binjiang) Blockchain and Data Security Research Institute, Hangzhou, China.
E-mail: \{xyin, chaoni\}@zju.edu.cn.}
\thanks{Shaohua Wang is with Central University of Finance and Economics, China. 
E-mail: davidshwang@ieee.org.}
\thanks{Chao Ni is the corresponding author.}
}

\IEEEtitleabstractindextext{%
\begin{abstract}
\justifying{
\xy
{
This paper proposes a pipeline for quantitatively evaluating interactive Large Language Models (LLMs) using publicly available datasets. 
We carry out an extensive technical evaluation of LLMs using Big-Vul covering four different common software vulnerability tasks. 
This evaluation assesses the multi-tasking capabilities of LLMs based on this dataset.
We find that the existing state-of-the-art approaches and pre-trained Language Models (LMs) are generally superior to LLMs in software vulnerability detection. 
However, in software vulnerability assessment and location, certain LLMs (e.g., CodeLlama and WizardCoder) have demonstrated superior performance compared to pre-trained LMs, and providing more contextual information can enhance the vulnerability assessment capabilities of LLMs.
Moreover, LLMs exhibit strong vulnerability description capabilities, but their tendency to produce excessive output significantly weakens their performance compared to pre-trained LMs. 
Overall, though LLMs perform well in some aspects, they still need improvement in understanding the subtle differences in code vulnerabilities and the ability to describe vulnerabilities to fully realize their potential.
Our evaluation pipeline provides valuable insights into the capabilities of LLMs in handling software vulnerabilities.
}
}

\end{abstract}

}

\maketitle

\IEEEdisplaynontitleabstractindextext

\IEEEpeerreviewmaketitle

\begin{figure}[htbp]
    \centering
    \vspace{0.3cm}
    \includegraphics[width=\linewidth]{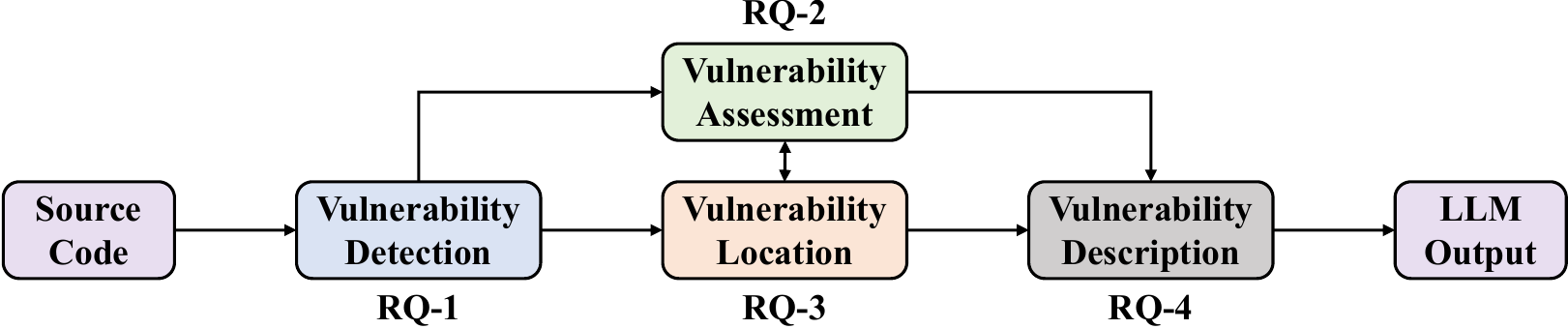}
    \caption{The relationship among software vulnerability analysis activities}
    \vspace{0.3cm}    
    \label{fig:relationship}
\end{figure}

\IEEEraisesectionheading{\section{Introduction}}
Software Vulnerabilities (SVs) can expose software systems to risk situations and eventually cause huge economic losses or even threaten people's lives.
Therefore, completing software vulnerabilities is an important task for software quality assurance (SQA).
Generally, there are many important software quality activities for software vulnerabilities such as SV detection, SV assessment, SV location, and SV description.
The relationship among the SQA activities is intricate and interdependent and can be illustrated in Fig.~\ref{fig:relationship}. 
SV detection serves as the initial phase, employing various tools and techniques to identify potential vulnerabilities within the software.
Once detected, the focus shifts to SV assessment, where the severity and potential impact of each vulnerability are meticulously evaluated. 
This critical evaluation informs the subsequent steps in the process.
SV location follows the assessment, pinpointing the exact areas within the software's code or architecture where vulnerabilities exist.
This step is crucial for precise remediation efforts and to prevent the recurrence of similar vulnerabilities in the future.
The intricacies of SV location feed into the comprehensive SV description, which encapsulates detailed information about each vulnerability, including its origin, characteristics, and potential exploits.
In essence, the synergy among SV detection, SV assessment, SV location, and SV description creates a robust pipeline for addressing software vulnerabilities comprehensively. 
This systematic approach not only enhances the overall quality of the software but also fortifies it against potential threats, thereby safeguarding against economic losses and potential harm to individuals. 
As a cornerstone of software quality assurance, the seamless integration of these activities underscores the importance of a proactive and thorough approach to managing software vulnerabilities in today's dynamic and interconnected digital landscape.

\begin{figure*}[!htbp]
\centering
\subfigure[LLMs' performance on different software vulnerability tasks \newline ($\ast$ refers to the results under fine-tuning setting)]{
\begin{minipage}[t]{0.495\linewidth}
\centering
\includegraphics[width=\linewidth]{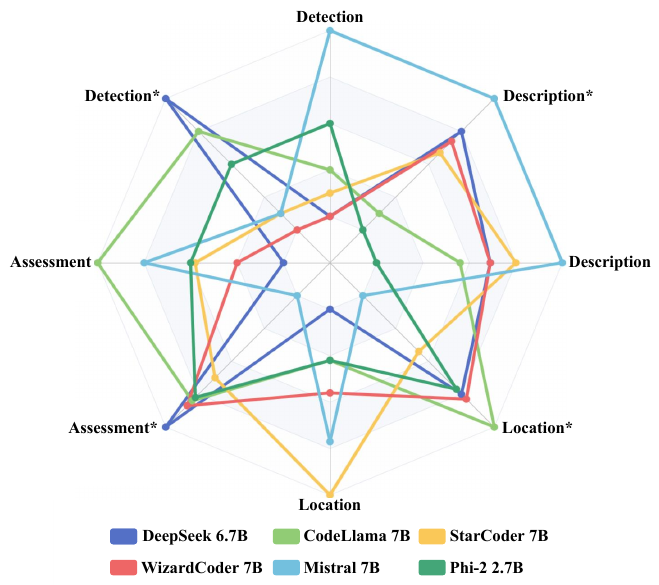}
\label{fig:radar1}
\end{minipage}%
}%
\subfigure[The impacts of parameter sizes on LLMs' performance across different software vulnerability tasks]{
\begin{minipage}[t]{0.495\linewidth}
\centering
\includegraphics[width=\linewidth]{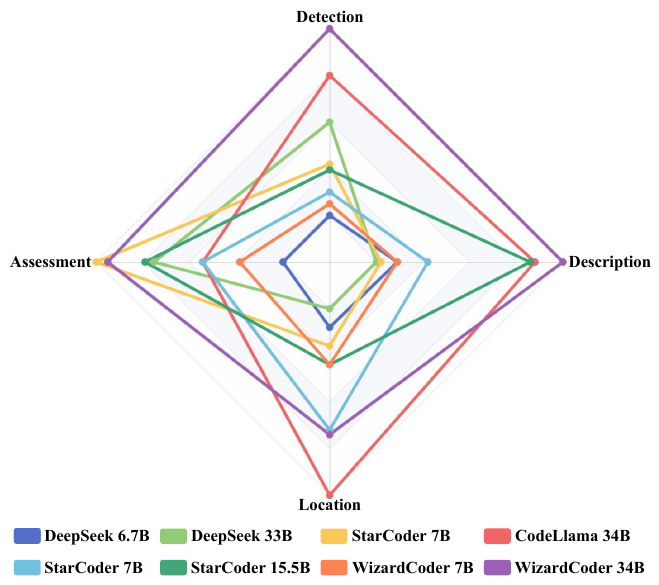}
\label{fig:radar2}
\end{minipage}%
}%
\centering
\caption{\xy{The capability comparison of LLMs with different parameter sizes on different software vulnerability tasks}}
\label{fig:radar}
\end{figure*}

Recently, Large Language Models (LLMs)~\cite{brown2020language} have been widely adopted since the advances in Natural Language Processing (NLP) which enable LLM to be well-trained with both billions of parameters and billions of training samples, consequently bringing a large performance improvement on tasks adopted by LLMs. 
LLMs can be easily used for a downstream task by being fine-tuned~\cite{radford2018improving} or being prompted~\cite{liu2023pre} since they are trained to be general and they can capture different knowledge from various domain data.
Fine-tuning is used to update model parameters for a particular downstream task by iterating the model on a specific dataset while prompting can be directly used by providing natural language descriptions or a few examples of the downstream task.
Compared to prompting, fine-tuning is expensive since it requires additional model training and has limited usage scenarios, especially in cases where sufficient training datasets are unavailable.

LLMs have demonstrated remarkable language comprehension and generation capabilities, and have been able to perform well on a variety of natural language processing tasks, such as text summarization~\cite{tang2023science}.
Given the outstanding performance of LLMs, there is a growing focus on exploring their potential in software engineering tasks and seeking new opportunities to address them.
Currently, as more and more LLMs designed for software engineering tasks are deployed~\cite{nijkamp2022codegen, zheng2023codegeex, wang2021codet5, wang2023codet5+, deepseek-coder, li2023starcoder, roziere2023code}, many research works focused on the application of LLMs in the software engineering domain~\cite{xia2023automated, xia2023keep, xia2022less, pan2023understanding, kang2023large}.
However, in the existing literature, adequate systematic reviews and surveys have been conducted on LLMs in areas such as generating high-quality code and high-coverage test cases~\cite{zan2023large, lemieux2023codamosa}, but a systematic review and evaluation of open-source LLMs in the field of software vulnerability is still missing.

In this paper, we focus on evaluating LLMs' performance in various software vulnerability (SV)-related tasks in few-shot and fine-tuning settings to obtain a basic, comprehensive, and better understanding of their multi-task ability, and we aim to answer the following research questions.

\begin{itemize}

    \item 
    \xy
    {
    \textbf{RQ-1: How do LLMs perform on vulnerability detection?}  
    Software Vulnerabilities (SVs) can expose software systems to risk situations and consequently software function failure.
    Therefore, detecting these SVs is an important task for software quality assurance. 
    We aim to explore the ability of LLMs on vulnerability detection as well as the performance difference compared with state-of-the-art approaches and pre-trained Language Models (LMs).
    }

    \item 
    \xy
    {
    \textbf{RQ-2: How do LLMs perform on vulnerability assessment?} 
    In practice, due to the limitation of SQA resources~\cite{khan2018review}, it is impossible to treat all detected SVs equally and fix all SVs simultaneously. 
    Thus, it is necessary to prioritize these detected software vulnerabilities for better treatment. 
    An effective solution to prioritize those SVs is to use one of the most widely known SV assessment frameworks CVSS (Common Vulnerability Scoring System)~\cite{le2021survey}, which characterizes SVs by considering three metric groups: Base, Temporal, and Environmental. 
    The metrics that are in the groups can be further used as the criterion for selecting serious SVs to fix early. 
    Therefore, we aim to explore the ability of LLMs to assess vulnerabilities and compare their performance with pre-trained LMs.
    }
    
    \item 
    \xy
    {
    \textbf{RQ-3: How do LLMs perform on vulnerability location?} 
    Identifying the precise location of vulnerabilities in software systems is of critical importance for mitigating risks and improving software quality. 
    The vulnerability location task involves pinpointing these weaknesses accurately and helps to narrow the scope for developers to fix problems. 
    Therefore, we aim to investigate LLMs' capability in effectively identifying the precise location of vulnerabilities in software systems, alongside evaluating their performance against state-of-the-art approaches and pre-trained LMs.
    }

    \item 
    \xy
    {
    \textbf{RQ-4: How do LLMs perform on vulnerability description?}
    The vulnerability description task focuses on conveying a detailed explanation of these identified issues in the source codes and helps participants to better understand the risk as well as its impacts. 
    Understanding the intricacies of vulnerabilities in software systems plays a pivotal role in alleviating risks and bolstering software quality. 
    The vulnerability description task focuses on conveying a detailed explanation of these identified issues in the source codes and helps participants to better understand the risk as well as its impacts. 
    Our goal is to evaluate LLMs' ability to effectively generate vulnerability descriptions within software systems and compare their performance with that of pre-trained LMs.
    }
    
\end{itemize}

To extensively and comprehensively analyze the LLMs' ability, we use a large-scale dataset containing real-world project vulnerabilities (named Big-Vul~\cite{fan2020ac}).
We carefully design experiments to discover the findings by answering four RQs.
The main contribution of our work is summarized as follows and takeaway findings are shown in Table~\ref{tab:takeaway}.
Eventually, we present the comparison of LLMs across four software vulnerability tasks under different settings, as well as the impact of varying model sizes on performance, as depicted in Fig.~\ref{fig:radar1} and Fig.~\ref{fig:radar2}.
In summary, the key contributions of this paper include:

\begin{itemize}[leftmargin=*]
    \item We extensively evaluate the performance of LLMs on different software vulnerability tasks and conduct an extensive comparison among LLMs and learning-based approaches to software vulnerability.
    \item We design four RQs to comprehensively understand LLMs from different dimensions, and provide detailed results with examples.
    \item We release our replication package for further study~\cite{replication}.
\end{itemize}

\begin{table*}[!ht]
    \centering
    \caption{\xy{Takeaways: Evaluating LLMs on Software Vulnerability}}
    \resizebox{\linewidth}{!}
    {
      \begin{tabular}{L{13em}|L{45em}}
      \toprule
      \textbf{Dimension} &\textbf{Finding or Guidance}\\
      \midrule
      \textbf{Vulnerability Detection} & 
      {\bf \circled{1}}. LLMs can detect vulnerabilities, but fine-tuned LLMs perform weaker than transformer-based approaches.
      Considering the computational resources and time costs of deploying LLMs, transformer-based approaches for vulnerability detection are a more efficient choice.
      {\bf \circled{2}}. After fine-tuning, the detection capability of LLMs has improved. 
      Larger models usually perform better, but performance can also be influenced by model design and pre-training data. 
      Therefore, fine-tuning the LLM on domain-specific data before using it as a vulnerability detector is necessary.
      {\bf \circled{3}}. In general, different LLMs complementing each other, while CodeLlama obtains better performance in terms of F1-score, Precision, and Recall.  \\
\midrule
       \textbf{Vulnerability Assessment} &
        {\bf \circled{4}}. Overall, fine-tuned code-related LLMs outperform pre-trained language models in vulnerability assessment. When resources permit, fine-tuning DeepSeek-Coder 6.7B for vulnerability assessment is optimal, as it outperforms the pre-trained language models across four metrics.
        {\bf \circled{5}}. LLMs have the capacity for assessment of vulnerability severity based on source code, and can be improved by providing more context information. \\
       \midrule
      \textbf{Vulnerability Location}& 
       {\bf \circled{6}}. Few-shot setting exposes LLM's limitations, and fine-tuning can greatly enhance the vulnerability location capabilities of LLMs.
       {\bf \circled{7}}. Fine-tuning code-related LLMs as vulnerability locators is beneficial, as they can outperform pre-trained language models in terms of F1-score, precision, and FPR. \\
      \midrule
        \textbf{Vulnerability Description}& 
         {\bf \circled{8}}. LLMs exhibit significantly weaker performance in generating vulnerability descriptions compared to pre-trained language models. Therefore, fine-tuning pre-trained language models for vulnerability detection is recommended.  \\
         \bottomrule
       \end{tabular}
       }
       \label{tab:takeaway}
\end{table*}

\section{Background and Related Work}

\subsection{Large Language Model}
{
Since the advancements in Natural Language Processing, Large Language Models (LLMs)~\cite{brown2020language} have seen widespread adoption due to their capacity to be effectively trained with billions of parameters and training samples, resulting in significant performance enhancements.
LLMs can readily be applied to downstream tasks through either fine-tuning~\cite{radford2018improving} or prompting~\cite{liu2023pre}.
Their versatility stems from being trained to possess a broad understanding, enabling them to capture diverse knowledge across various domains.
Fine-tuning involves updating the model parameters specifically for a given downstream task through iterative training on a specific dataset. 
In contrast, prompting allows for direct utilization by providing natural language descriptions or a few examples of the downstream task.
Compared to prompting, fine-tuning is resource-intensive as it necessitates additional model training and is applicable in limited scenarios, particularly when adequate training datasets are unavailable.
}

LLMs are usually built on the transformer architecture~\cite{vaswani2017attention} and can be classified into three types of architectures: encoder-only, encoder-decoder, and decoder-only. 
Encoder-only (e.g., CodeBERT~\cite{feng2020codebert}, GraphCodeBERT~\cite{guo2020graphcodebert}, and UniXcoder~\cite{guo2022unixcoder}) and Encoder-Decoder (e.g., PLBART~\cite{ahmad2021unified}, CodeT5~\cite{wang2021codet5}, and CodeT5+~\cite{wang2023codet5+}) models are trained using Masked Language Modeling (MLM) or Masked Span Prediction (MSP) objective, respectively, where a small portion (e.g., 15\%) of the tokens are replaced with either masked tokens or masked span tokens, LLMs are trained to recover the masked tokens.
These models are trained as general ones on the code-related data and then are fine-tuned for the downstream tasks to achieve superior performance. 
Decoder-only models also attract a small portion of people's attention and they are trained by using Causal Language Modeling objectives to predict the probability of the next token given all previous tokens. 
GPT~\cite{radford2018improving} and its variants are the most representative models, which bring the large language models into practical usage.

Recently, the ChatGPT model attracts the widest attention from the world, which is the successor of the large language model InstructGPT~\cite{ouyang2022training} with a dialog interface that is fine-tuned using the Reinforcement Learning with Human Feedback (RLHF) approach~\cite{christiano2017deep,ouyang2022training,ziegler2019fine}.
RLHF initially fine-tunes the base model using a small dataset of prompts as input and the desired output, typically human-written, to refine its performance.
Subsequently, a reward model is trained on a larger set of prompts by sampling outputs generated by the fine-tuned model. 
These outputs are then reordered by human labelers to provide feedback for training the reward model.
Reinforcement learning~\cite{schulman2017proximal} is then used to calculate rewards for each output generated based on the reward model, updating LLM parameters accordingly. 
With fine-tuning and alignment with human preferences, LLMs better understand input prompts and instructions, enhancing performance across various tasks~\cite{bang2023multitask,ouyang2022training}.

\xy
{
The application of LLMs in software engineering has seen a surge, with models like ChatGPT being employed for various tasks 
(e.g., code review, code generation, and vulnerability detection).
Although some works use LLMs for vulnerability tasks~\cite{zhou2024large,fu2023chatgpt}, our work differs from these previous studies in the following aspects.
\textbf{(1) Closed-source ChatGPT vs. Open-source LLMs:} They only explore the capabilities of the closed-source ChatGPT in vulnerability tasks, whereas we investigate the abilities of both open-source code-related LLMs and general LLMs in these tasks.
\textbf{(2) Prompts vs. Few-shot and Fine-tuning Settings:} 
They focus solely on the performance of LLMs using prompts, which introduces randomness and hinders the reproducibility of their findings.
In contrast, we examine the capabilities of LLMs under both few-shot and fine-tuning settings, providing the source code and corresponding model files to ensure the reproducibility of our experimental results.
}

\subsection{Software Vulnerability}
\xy
{
Software Vulnerabilities (SVs) can expose software systems to risk situations and consequently make the software under cyber-attacks, eventually causing huge economic losses and even threatening people's lives.
Therefore, vulnerability databases have been created to document and analyze publicly known security vulnerabilities.
For example, Common Vulnerabilities and Exposures (CVE)~\cite{cve,bhandari2021cvefixes} and SecurityFocus~\cite{securityfocus} are two well-known vulnerability databases.
Besides, Common Weakness Enumeration (CWE) defines the common software weaknesses of individual vulnerabilities, which are often referred to as vulnerability types of CVEs. 
To better address these vulnerabilities, researchers have proposed many approaches for understanding the effects of software vulnerabilities, including SV detection~\cite{zhou2019devign,fu2022linevul,cao2022mvd,li2021sysevr,cheng2022path,wu2022vulcnn,li2018vuldeepecker,hin2022linevd,zhan2021atvhunter,chakraborty2021deep,ni2023distinguishing,steenhoek2023empirical,steenhoek2024dataflow}, SV assessment~\cite{ni2023fva,feutrill2018effect,le2021survey,spanos2018multi,le2021deepcva}, SV location~\cite{li2021vuldeelocator,li2021vulnerability,ni2022best}, SV repair~\cite{ni2022defect,zhang2022program,chen2019sequencer,zhu2021syntax} as well as SV description~\cite{sun2021generating,guo2022detecting,guo2021key,guo2020predicting}.
Many novel technologies are adopted to promote the progress of software vulnerability management, including software analysis~\cite{fan2019smoke,li2020pca}, machine learning~\cite{zhou2019devign,hin2022linevd}, and deep learning~\cite{ni2023fva,li2021vulnerability}, especially LLMs~\cite{guo2022detecting,guo2021key}.
}

\section{Experimental Design}

\xy
{
In this section, we present our studied dataset, our studied LLMs, the techniques for fine-tuning, the prompt engineering, the baseline approaches, the evaluation metrics, and the experiment settings.
}

\subsection{Studied Dataset}
\label{lab:dataset}

\xy
{
We adopt the widely used dataset (named Big-Vul) provided by Fan et al.~\cite{fan2020ac} by considering the following reasons. 
The most important one is to satisfy the distinct characteristics of the real world as well as the diversity in the dataset, which is suggested by previous works~\cite{hin2022linevd,chakraborty2021deep}.
Big-Vul, to the best of our knowledge, is the most large-scale vulnerability dataset with diverse information about the vulnerabilities, which are collected from practical projects and these vulnerabilities are recorded in the Common Vulnerabilities and Exposures (CVE)\footnote{https://cve.mitre.org/}.
The second one is to compare fairly with existing state-of-the-art (SOTA) approaches (e.g., LineVul, Devign, and SVulD).
}

\xy
{
Big-Vul totally contains 3,754 code vulnerabilities collected from 348 open-source projects spanning 91 different vulnerability types from 2002 to 2019.
It has 188,636 C/C++ functions with a vulnerable ratio of 5.7\% (i.e., 10,900 vulnerability functions).
The authors linked the code changes with CVEs and their descriptive information to enable a deeper analysis of the vulnerabilities.
}

\xy
{
We follow the same strategy to build the training data, validation data, and testing data from the original dataset with previous work does~\cite{fu2022linevul, ni2022defect, ni2023distinguishing}. 
Specifically, 80\% of functions are treated as training data, 10\% of functions are treated as validation data, and the left 10\% of functions are treated as testing data. 
We also keep the distribution as same as the original ones in training, validation, and testing data.
Notice that we undersample the non-vulnerable functions to produce approximately balanced training data at the function level, while the validation and testing data remain in the original imbalanced ratio.
To clean and normalize the dataset, we remove empty lines, leading and trailing spaces in each line, as well as comments from the source code.
Finally, the split dataset is used for evaluation and the statistics are shown in Table~\ref{tab:dataset}.
}

\begin{table}[htbp]
  \vspace{0.3cm}
  \centering
  \caption{\xy{Statistic of the studied dataset}}
  \resizebox{\linewidth}{!}{
  \begin{threeparttable}
    \begin{tabular}{lrrrr}
    \toprule
    \textbf{Datasets} & \textbf{\# Vul.} & \textbf{\# Non-Vul.} & \textbf{\# Total} &\textbf{\% Vul.: Non-Vul.} \\
    \midrule
    Original Big-Vul & 10,900  & 177,736  & 188,636 & 0.061 \\
    Filtered Big-Vul & 5,260 & 96,308 & 101,568 & 0.055 \\
    \midrule
    Training & 8,720 & 8,720  & 17,440  & 1 \\
    Validation & 1,090 & 17,774 & 18,864 & 0.061 \\
    Testing & 1,090 & 17,774 & 18,864 & 0.061 \\
    \bottomrule
    \end{tabular}%
  $^\ast$We undersample the non-vulnerable functions to produce approximately balanced training data.
  \end{threeparttable}
  }
  \label{tab:dataset}%
\end{table}%

\subsection{Studied LLMs}
\label{lab:llms}

The general LLMs are pre-trained on textual data, including natural language and code, and can be used for a variety of tasks. 
In contrast, code-related LLMs are specifically pre-trained to automate code-related tasks. 
Due to the empirical nature of this work, we are interested in assessing the effectiveness of both LLM categories in vulnerability tasks. 
For the code-related LLMs, we select the top four models released recently (in 2023), namely DeepSeek-Coder~\cite{deepseek-coder}, CodeLlama~\cite{roziere2023code}, StarCoder~\cite{li2023starcoder}, and WizardCoder~\cite{luo2023wizardcoder}. 
For the general LLMs, we select the top two models, resulting in the selection of Mistral~\cite{jiang2023mistral}, and Phi-2~\cite{phi}. 
For the few-shot setting, we select the models with no more than 34B parameters from the Hugging Face Open LLM Leaderboard~\cite{huggingface2023llm}, as for the fine-tuning setting, we select the models with 7B parameters or less.
The constraint on the number of parameters is imposed by our computing resources (i.e., 192GB RAM, 10 × NVIDIA RTX 3090 GPU).
Table~\ref{tab:llm} summarizes the characteristics of the studied LLMs, we briefly introduce these LLMs to make our paper self-contained.

\begin{table}[htbp]
  \centering
  \caption{Overview of the studied LLMs}
  \resizebox{\linewidth}{!}
  {
    \begin{threeparttable}
    \begin{tabular}{l|cccc|cc}
    \toprule
    \multicolumn{1}{l|}{\multirow{2}[2]{*}{\textbf{Models}}} & \multicolumn{4}{c|}{\textbf{Code-related LLMs}} & \multicolumn{2}{c}{\textbf{General LLMs}} \\
    \cmidrule{2-7} 
    \multicolumn{1}{l|}{} & \textbf{DeepSeek-Coder} & \textbf{CodeLlama} & \textbf{StarCoder} & \textbf{WizardCoder} & \textbf{Mistral} & \textbf{Phi-2} \\
    \midrule
    \textbf{Fine-Tuning} & 6.7B & 7B & 7B & 7B & 7B & 2.7B \\
    \textbf{Few-Shot} & 6.7B \& 33B & 7B \& 34B & 7B \& 34B & 7B \& 15.5B & 7B & 2.7B \\
    \midrule
    \textbf{Release Date} & Nov'23 & Aug'23 & May'23 & June'23 & Sep'23 & Dec'23  \\
    \bottomrule
    \end{tabular}%
    \end{threeparttable}
    }
  \label{tab:llm}%
\end{table}%


\noindent
\textbf{Group 1: Code-related LLMs.}
\textbf{DeepSeek-Coder} developed by DeepSeek AI~\cite{deepseek-coder} is composed of a series of code language models, each trained from scratch on 2T tokens, with a composition of 87\% code and 13\% natural language in both English and Chinese. 
They provide various sizes of the code model, ranging from 1B to 33B versions. 
Each model is pre-trained on project-level code corpus by employing a window size of 16K and an extra fill-in-the-blank task, to support project-level code completion and infilling. 
For coding capabilities, DeepSeek-Coder achieves state-of-the-art performance among open-source code models on multiple programming languages and various benchmarks.

\textbf{CodeLlama} proposed by Rozière et al.~\cite{roziere2023code} is a set of large pre-trained language models for code built on Llama 2.
They achieve state-of-the-art performance among open models on code tasks, provide infilling capabilities, support large input contexts, and demonstrate zero-shot instruction following for programming problems. 
CodeLlama is created by further training Llama 2 using increased sampling of code data. 
As with Llama 2, the authors applied extensive safety mitigations to the fine-tuned CodeLlama versions.

\textbf{StarCoder} proposed by Li et al.~\cite{li2023starcoder} is a large pre-trained language model specifically designed for code. 
It was pre-trained on a large amount of code data to acquire programming knowledge and trained on permissive data from GitHub, including over 80 programming languages, Git commits, GitHub issues, and Jupyter notebooks. 
StarCoder can perform code editing tasks, understand natural language prompts, and generate code that conforms to APIs. 
StarCoder represents the advancement of applying large language models in programming.

\textbf{WizardCoder} proposed by Luo et al.~\cite{luo2023wizardcoder} is a large pre-trained language model that empowers Code LLMs with complex instruction fine-tuning, by adapting the Evol-Instruct method to
the domain of code. 
Through comprehensive experiments on four prominent code generation benchmarks, namely HumanEval, HumanEval+, MBPP, and DS-1000, the authors unveil the exceptional capabilities of their model. 
It surpasses all other open-source Code LLMs by a substantial margin. 
Moreover, WizardCoder even outperforms the largest closed LLMs, Anthropic’s Claude and Google’s Bard, on HumanEval and HumanEval+.

\noindent
\textbf{Group 2: General LLMs.}
\textbf{Mistral} is a 7-billion-parameter language model released by Mistral AI~\cite{jiang2023mistral}. 
Mistral 7B is a carefully designed language model that provides both efficiency and high performance to enable real-world applications.
Due to its efficiency improvements, the model is suitable for real-time applications where quick responses are essential. 
At the time of its release, Mistral 7B outperformed the best open source 13B model (Llama 2) in all evaluated benchmarks.

\textbf{Phi-2} proposed by Microsoft~\cite{phi} packed with 2.7 billion parameters. 
It is designed to make machines think more like humans and do it safely. 
Phi-2 is not just about numbers; it is about a smarter, safer way for computers to understand and interact with the world.
Phi-2 stands out because it is been taught with a mix of new language data and careful checks to make sure it acts right.
It is built to do many things like writing, summarizing texts, and coding, but with better common sense and understanding than its earlier version, Phi-1.5.
Phi-2's evaluation demonstrates its proficiency over larger models in aggregated benchmarks, emphasizing the potential of smaller models to achieve comparable or superior performance to their larger counterparts. This is particularly evident in its comparison with Google Gemini Nano 2, where Phi-2 outshines despite its smaller size.

\subsection{Model Fine-Tuning}
\label{sec:fine-tuning}
\begin{figure}[!htbp]
\centering
\subfigure[Discriminative Fine-Tuning]{
\begin{minipage}[t]{0.485\linewidth}
\centering
\includegraphics[width=0.995\linewidth]{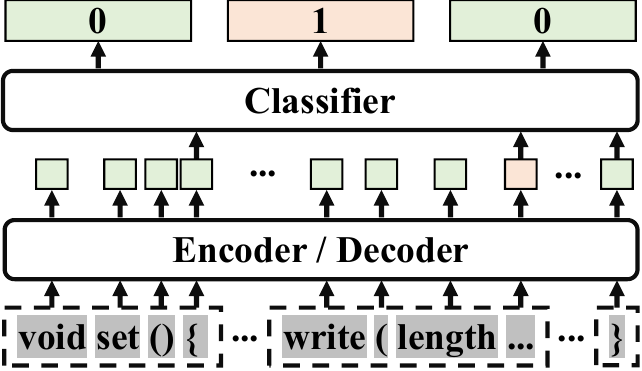}
\label{fig:discriminative}
\end{minipage}%
}%
\subfigure[Generative Fine-Tuning]{
\begin{minipage}[t]{0.485\linewidth}
\centering
\includegraphics[width=\linewidth]{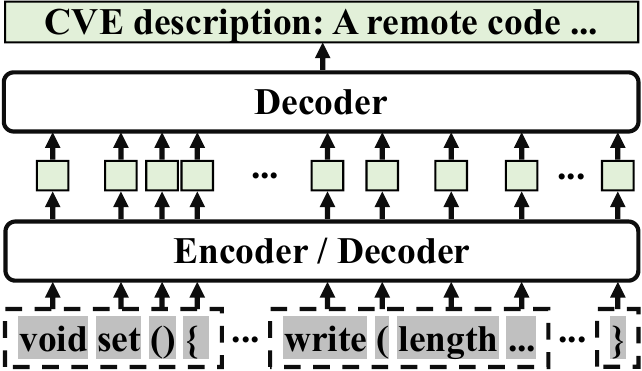}
\label{fig:generative}
\end{minipage}%
}%
\centering
\caption{\xy{Fine-tuning LLMs for software vulnerability tasks}}
\label{fig:overview}
\end{figure}

\xy
{
The four software vulnerability tasks can be categorized into two types: discriminative task (i.e., software vulnerability detection, software vulnerability assessment, and software vulnerability location) and generative task (i.e., software vulnerability description).
Therefore, fine-tuning LLMs for software vulnerability tasks can be undertaken through both discriminative and generative methods, each method specifically designed to make LLMs aligned with the task.
In particular, we treat the discriminative tasks as binary classification, while treating the generative task as generation one.
The architectures for the two paradigms are presented in Fig.~\ref{fig:overview}.
}

\noindent
\xy
{
\textbf{Discriminative Fine-Tuning.}
For vulnerability detection and vulnerability assessment, we utilize the \textit{``AutoModelForSequenceClassification''} class provided by the Transformers library to implement discriminative fine-tuning. \textit{``AutoModelForSequenceClassification''} is a generic model class that will be instantiated as one of the sequence classification model classes of the library when created with the \textit{``AutoModelForSequenceClassification.from\_pretrained(model\_name\_or\_path)''} class method.
}

\xy
{
For vulnerability location, we follow previous works~\cite{yang2024large,zhang2024empirical} that use LLMs to classify individual code lines as either vulnerable or non-vulnerable. 
For a token sequence $T=\{t_1, t_2,...,t_n\}$ of the function, the model’s decoder component, denoted as $M$, processes $T$ to yield a sequence of output vectors: $O=M(T)=\{o_1, o_2,..., o_L\}$, where $O$ represents the output tensor with dimensions $L \times H$, $L$ signifies the sequence length, and $H$ denotes the hidden dimension size.
During the process, the contextual information is captured by the masked self-attention mechanisms in the decoder of LLMs, where masked self-attention limits the sight to the preceding part of tokens.
Each output vector $o_i$ that represents the last token of one line is subsequently associated with a label (i.e., 0 or 1).
The optimization process employs the binary cross-entropy as the loss function.
}

\begin{table*}[!th]
    \centering
    \caption{The task descriptions and indicators for different software vulnerability tasks}
    
    \resizebox{1.0\linewidth}{!}
    {
      \begin{tabular}{l|l|l}
      \toprule
      \textbf{Dimension} & \textbf{Task Description} & \textbf{Indicator} \\
      \midrule
      \textbf{Vulnerability Detection} & 
      \tabincell{l}{
      If this C code snippet has vulnerabilities, output Yes; otherwise, output No. \\
      } 
      & // Detection \\
\midrule
       \textbf{Vulnerability Assessment} & \tabincell{l}{
       Provide a qualitative severity ratings of CVSS v2.0 for the vulnerable C code snippet. \\
       } 
       & // Assessment \\
       \midrule
      \textbf{Vulnerability Location}& \tabincell{l}{
      Provide a vulnerability location result for the vulnerable C code snippet.
      } 
      & // Location \\
       \midrule
        \textbf{Vulnerability Description}& \tabincell{l}{
        Provide a CVE description for the vulnerable C code snippet.
        }  
        & // Description \\
         \bottomrule
       \end{tabular}
       }
       \label{tab:prompt}
\end{table*}

\noindent
\xy
{
\textbf{Generative Fine-Tuning.}
Generative fine-tuning aims to equip LLMs with the ability to perform Sequence-to-Sequence (Seq2Seq) tasks. 
Specifically, this involves inputting vulnerable code and generating the corresponding CVE descriptions related to the vulnerabilities.
To calculate the loss during fine-tuning, we utilize the cross-entropy loss function, which is commonly used for Seq2Seq tasks.  
In this context, the loss measures the difference between the generated output sequence and the target sequence.  
}

\subsection{Prompt Engineering}
\label{lab:prompt}
\begin{figure}[htp]
    \centering
    \includegraphics[width=\linewidth]{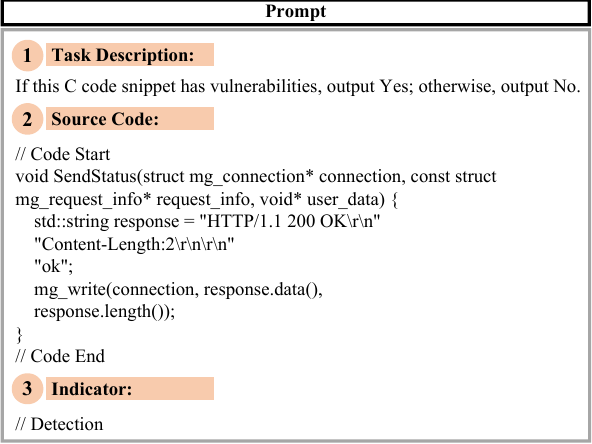}
    \caption{The prompt contains three pieces of information: (1) task description, (2) source code, and (3) indicator}
    \label{fig:prompt}
\end{figure}

\xy
{
For few-shot setting, we follow the prompt similar to those used in the artifacts, papers, or technical reports associated with each corresponding model~\cite{nijkamp2022codegen, li2023starcoder, roziere2023code}, where each prompt contains three pieces of information: (1)
task description, (2) source code, and (3) indicator. 
Using the software vulnerability detection task as an example, the prompt utilized for LLM consists of three crucial components, as depicted in Fig.~\ref{fig:prompt}:
}
\begin{itemize}[leftmargin=*]
\item \textbf{Task Description} (marked as \ding{172}). We provide LLM with the description constructed as \texttt{``If this C code snippet has vulnerabilities, output Yes; otherwise, output No''}. 
The task description used in the SV detection task varies based on the source programming language we employ.
\item \textbf{Source Code} (marked as \ding{173}). 
We provide LLM with the code wrapped in \texttt{``// Code Start''} and \texttt{``// Code End''}
Since we illustrate an example in C, we use the C comment format of \texttt{``//''} as a prefix for the description. 
We also employ different comment prefixes based on the programming language of the code.
\item \textbf{Indicator} (marked as \ding{174}). 
We instruct LLM to think about the results.
In this paper, we follow the best practice in previous work~\cite{xia2023automated} and adopt the same prompt named \texttt{``// Detection''}.
\end{itemize}

Depending on the specific software vulnerability tasks, the task descriptions and indicators in the prompts may vary. 
The task descriptions and indicators for different software vulnerability tasks are presented in Table~\ref{tab:prompt}.

\subsection{Baselines}
\xy
{
To comprehensively compare the performance of LLMs with existing approaches, in this study, we consider the various pre-trained Language Models (LMs). 
As shown in Table~\ref{tab:studied_lms}, these models have fewer than 220 million parameters and can be categorized into two categories: encoder-only LMs and encoder-decoder LMs.
Encoder-only LMs (i.e., CodeBERT~\cite{feng2020codebert}, GraphCodeBERT~\cite{guo2020graphcodebert}, and UniXcoder~\cite{guo2022unixcoder}) contain only the encoder component of a Transformer. 
They are designed for learning data representations and trained using the Masked Language Modeling (MLM) objective. 
Encoder-decoder LMs (i.e., PLBART~\cite{ahmad2021unified}, CodeT5~\cite{wang2021codet5}, and CodeT5+~\cite{wang2023codet5+}) have been proposed for sequence-to-sequence tasks.
They are trained to recover the correct output sequence given the original input, often through span prediction tasks where random spans are replaced with artificial tokens. 
Recently, researchers have combined MLM with generative models for bidirectional and autoregressive text generation or infilling~\cite{aghajanyan2022cm3}.
All these LMs can potentially be used for our tasks, so we evaluate these LMs.
}

\begin{table}[htbp]
  \centering
  \vspace{0.2cm}
  \caption{\xy{Overview of the studied LMs}}
  \resizebox{\linewidth}{!}
  {
    \begin{threeparttable}
    \begin{tabular}{ccc|ccc}
    \toprule
    \textbf{Models} & \textbf{\# Para.} & \textbf{Model Type} & \textbf{Models} & \textbf{\# Para.} & \textbf{Model Type} \\
    \midrule
    CodeBERT & 125M & Encoder-only LM & PLBART & 140M & Encoder-decoder LM \\
    GraphCodeBERT & 125M & Encoder-only LM & CodeT5 & 220M & Encoder-decoder LM  \\
    UniXcoder & 125M & Encoder-only LM & CodeT5+ & 220M & Encoder-decoder LM \\
    \bottomrule
    \end{tabular}%
    $^\ast$For UniXcoder, we use encoder-only mode.
    \end{threeparttable}
  }
  \vspace{0.2cm}
  \label{tab:studied_lms}%
\end{table}%

\xy
{
For vulnerability location, we also consider Devign~\cite{zhou2019devign}, Reveal~\cite{chakraborty2021deep}, IVDetect~\cite{li2021vulnerability}, and LineVul~\cite{fu2022linevul} as baselines.
In addressing vulnerability detection, we also include SVulD~\cite{ni2023distinguishing} in addition to the aforementioned approaches. 
We briefly introduce them as follows.
}

\textbf{Devign} proposed by Zhou et al.~\cite{zhou2019devign} is a general graph neural network-based model for graph-level classification through learning on a rich set of code semantic representations including AST, CFG, DFG, and code sequences.
It uses a novel $Conv$ module to efficiently extract useful features in the learned rich node representations for graph-level classification.

{\textbf{Reveal}} proposed by Chakraborty et al.~\cite{chakraborty2021deep} contains two main phases. 
In the feature extraction phase, it translates code into a graph embedding, and in the training phase, it trains a representation learner on the extracted features to obtain a model that can distinguish the vulnerable functions from non-vulnerable ones.

{\textbf{IVDetect}} proposed by Li et  al.~\cite{li2021vulnerability} contains the coarse-grained vulnerability detection component and fine-grained interpretation component.
In particular, IVDetect represents source code in the form of a program dependence graph (PDG) and treats the vulnerability detection problem as graph-based classification via graph convolution network with feature attention.
As for interpretation, IVDetect adopts a GNNExplainer to provide fine-grained interpretations that include the sub-graph in PDG with crucial statements that are relevant to the detected vulnerability.

\textbf{LineVul} proposed by Fu et al.~\cite{fu2022linevul} is a Transformer-based line-level vulnerability prediction approach. 
LineVul leverages BERT architecture with self-attention layers which can capture long-term dependencies within a long sequence. 
Besides, benefiting from the large-scale pre-trained model, LineVul can intrinsically capture more lexical and logical semantics for the given code input.
Moreover, LineVul adopts the attention mechanism of BERT architecture to locate the vulnerable lines for finer-grained detection.

\textbf{SVulD} proposed by Ni et al.~\cite{ni2023distinguishing} is a function-level subtle semantic embedding for vulnerability detection along with heuristic explanations.
Particularly, SVulD adopts contrastive learning to train the UniXcoder semantic embedding model for learning distinguishing semantic representation of functions regardless of their lexically similar information.


\subsection{Evaluation Metrics}

\xy
{
For considered software vulnerability-related tasks, we will perform evaluations using the widely adopted performance metrics. 
More precisely, to evaluate the effectiveness of LLMs on vulnerability detection and vulnerability assessment, we consider the following four metrics: F1-score, Recall, Precision, and Accuracy.
Additionally, for vulnerability location, besides the four aforementioned metrics, we also consider the Top-k Accuracy and FPR metrics.
For vulnerability description, we use \textit{Rouge-1}, \textit{Rouge-2}, and \textit{Rouge-L} metrics.
}






\subsection{Implementation}
\label{sec:implementation}
We develop the generation pipeline in Python, utilizing PyTorch~\cite{pytorch} implementations of DeepSeek Coder, CodeLlama, StarCoder, WizardCoder, Mistral, and Phi-2. 
We use the Huggingface~\cite{huggingface} to load the model weights and generate outputs.
We also adhere to the best-practice guide~\cite{shieh2023best} for each prompt.
For the fine-tuning setting, we select the models with 7B parameters or less, and for the few-shot setting, we use models with fewer than 34B parameters.
To directly compare the fine-tuning setting with the few-shot setting, we employ models with the same parameter in both settings (i.e., DeepSeek-Coder 6.7B, CodeLlama 7B, StarCoder 7B, WizardCoder 7B, Mistral 7B, and Phi-2 2.7B).
The constraint on the number of parameters is imposed by our computing resources. 
Table~\ref{tab:llm} summarizes the characteristics of the studied LLMs.
Furthermore, considering the limitation of LLM's conversation windows, we manually select three examples for the few-shot setting from the training data.
Regarding baselines (i.e., pre-trained LMs, Reveal, IVDetect, Devign, LineVul, and SVulD), we utilize their publicly available source code and perform fine-tuning with the default parameters provided in their original code. 
Considering Devign's code is not publicly available, we make every effort to replicate its functionality and achieve similar results on the original paper's dataset.
All these models are implemented using the PyTorch~\cite{pytorch} framework.
The evaluation is conducted on a 16-core workstation equipped with an Intel(R) Xeon(R) Gold 6226R CPU @ 2.90Ghz, 192GB RAM, and 10 × NVIDIA RTX 3090 GPU, running Ubuntu 20.04.1 LTS.

\section{Experimental results}

This section presents the experimental results by evaluating LLMs performances on the widely used comprehensive dataset (i.e., Big-Vul~\cite{fan2020ac}) covering four SV-related tasks.

\begin{table}[htbp]
  \centering
  \caption{\xy{The comparison between LLMs and eleven baselines on software vulnerability detection (RQ1)}}
  \resizebox{\linewidth}{!}
  {
    \begin{tabular}{lcccc}
    \toprule
    \textbf{Methods} & \textbf{F1-score} & \textbf{Recall} & \textbf{Precision} & \textbf{Accuracy} \\
    \midrule
    Devign & 0.200 & 0.660 & 0.118  & 0.726  \\
    {Reveal} & 0.232  & 0.354  & 0.172  & 0.811  \\
    {IVDetect} & 0.231  & 0.540  & 0.148  & 0.815  \\
    LineVul & 0.272  & 0.620  & 0.174  & 0.828  \\
    SVulD & \textbf{0.336} & 0.414 & \textbf{0.282} & \textbf{0.915} \\
    CodeBERT & 0.270  & 0.608  & 0.173  & 0.830  \\
    GraphCodeBERT & 0.246  & 0.721  & 0.148  & 0.771  \\
    UniXcoder & 0.256  & 0.787  & 0.153  & 0.764  \\
    PLBART & 0.255  & 0.692  & 0.157  & 0.791  \\
    CodeT5 & 0.237  & 0.759  & 0.141  & 0.748  \\
    CodeT5+ & 0.218  & 0.508  & 0.139  & 0.812  \\
    \midrule
    \rowcolor{lightgray}\textbf{Fine-Tuning Setting} &&&& \\
    DeepSeek-Coder 6.7B & 0.270 & 0.627 & 0.172 & 0.824 \\
    CodeLlama 7B & 0.259 & \textbf{0.806} & 0.154 & 0.761 \\
    StarCoder 7B & 0.220 & 0.607 & 0.135 & 0.778 \\
    WizardCoder 7B & 0.214 & 0.365 & 0.151 & 0.861 \\
    Mistral & 0.220 & 0.607 & 0.135 & 0.778 \\
    Phi-2 & 0.241 & 0.557 & 0.154 & 0.818 \\
    \midrule
     \rowcolor{lightgray} \textbf{Few-Shot Setting} &&&&\\
    DeepSeek-Coder 6.7B & 0.084 & 0.156 & 0.057 & 0.823 \\
    DeepSeek-Coder 33B & 0.107 & 0.688 & 0.058 & 0.404 \\
    CodeLlama 7B & 0.098 & 0.449 & 0.055 & 0.570 \\
    CodeLlama 34B & 0.117 & 0.281 & 0.074 & 0.781 \\
    StarCoder 7B & 0.094 & 0.443 & 0.053 & 0.560 \\
    StarCoder 15.5B & 0.097 & 0.557 & 0.053 & 0.463 \\
    WizardCoder 7B& 0.086 & 0.380 & 0.049 & 0.583 \\
    WizardCoder 34B & 0.128 & 0.559 & 0.072 & 0.607 \\
    Mistral & 0.126 & 0.401 & 0.074 & 0.711 \\
    Phi-2 & 0.099 & 0.563 & 0.054 & 0.471 \\
    \bottomrule
    \end{tabular}%
    }
    
  \label{tab:rq1_llm_vs_sota}%
\end{table}%

\begin{table*}[htbp]
  \centering
  \caption{\xy{The software vulnerability detection comparison on Top-10 CWEs among fine-tuned LLMs (RQ1)}}
  \resizebox{\linewidth}{!}
  {
    \begin{tabular}{|lcc|cccccc|cccccc|}
    \toprule
    \multirow{2.2}[1]{*}{\textbf{CWE Type}} & \multirow{2.2}[1]{*}{\textbf{\# Total}} & \multirow{2.2}[1]{*}{\textbf{\# Vul.}} & \multicolumn{1}{l|}{\textbf{DeepSeek-Coder}} & \multicolumn{1}{l|}{\textbf{CodeLlama}} & \multicolumn{1}{l|}{\textbf{StarCoder}} & \multicolumn{1}{l|}{\textbf{WizardCoder}} & \multicolumn{1}{l|}{\textbf{Mistral}} & \multicolumn{1}{l|}{\textbf{Phi-2}} & \multicolumn{1}{l|}{\textbf{DeepSeek-Coder}} & \multicolumn{1}{l|}{\textbf{CodeLlama}} & \multicolumn{1}{l|}{\textbf{StarCoder}} & \multicolumn{1}{l|}{\textbf{WizardCoder}} & \multicolumn{1}{l|}{\textbf{Mistral}} & \multicolumn{1}{l|}{\textbf{Phi-2}} \\
    \cmidrule{4-15}      &   &   & \multicolumn{6}{c|}{\textbf{F1-score}} & \multicolumn{6}{c|}{\textbf{Precision}} \\
    \midrule
    CWE-119 & 1549 & 128 & \textbf{0.321} & 0.309 & 0.316 & 0.269 & 0.258 & 0.281 & \textbf{0.223} & 0.197 & 0.212 & 0.215 & 0.181 & 0.197 \\
    CWE-20 & 1082 & 80 & 0.269 & \textbf{0.273} & 0.229 & 0.216 & 0.145 & 0.269 & 0.173 & 0.163 & 0.141 & 0.165 & 0.096 & \textbf{0.175} \\
    CWE-264 & 800 & 64 & \textbf{0.486} & 0.468 & 0.337 & 0.348 & 0.356 & 0.477 & 0.357 & 0.316 & 0.232 & 0.308 & 0.269 & \textbf{0.361} \\
    CWE-399 & 697 & 35 & \textbf{0.355} & 0.286 & 0.209 & 0.274 & 0.227 & 0.306 & \textbf{0.227} & 0.169 & 0.125 & 0.191 & 0.143 & 0.196 \\
    CWE-125 & 582 & 29 & 0.233 & \textbf{0.267} & 0.213 & 0.195 & 0.179 & 0.180 & 0.145 & \textbf{0.156} & 0.129 & 0.128 & 0.108 & 0.109 \\
    CWE-200 & 573 & 27 & \textbf{0.269} & 0.261 & 0.241 & 0.180 & 0.162 & 0.229 & \textbf{0.182} & 0.159 & 0.151 & 0.132 & 0.106 & 0.152 \\
    CWE-189 & 442 & 21 & 0.235 & 0.208 & 0.255 & 0.273 & 0.178 & \textbf{0.293} & 0.145 & 0.119 & 0.151 & 0.180 & 0.108 & \textbf{0.182} \\
    CWE-362 & 413 & 16 & 0.031 & \textbf{0.086} & 0.075 & 0.050 & 0.026 & 0.032 & 0.017 & \textbf{0.045} & 0.040 & 0.029 & 0.014 & 0.018 \\
    CWE-416 & 406 & 12 & \textbf{0.193} & 0.178 & 0.148 & 0.145 & 0.146 & 0.141 & \textbf{0.113} & 0.101 & 0.083 & 0.093 & 0.086 & 0.082 \\
    CWE-476 & 367 & 11 & 0.091 & \textbf{0.109} & 0.053 & 0.057 & 0.037 & 0.019 & 0.049 & \textbf{0.057} & 0.028 & 0.032 & 0.020 & 0.010 \\
    \midrule
    \multirow{2.2}[1]{*}{\textbf{CWE Type}} & \multirow{2.2}[1]{*}{\textbf{\# Total}} & \multirow{2.2}[1]{*}{\textbf{\# Vul.}} & \multicolumn{1}{l|}{\textbf{DeepSeek-Coder}} & \multicolumn{1}{l|}{\textbf{CodeLlama}} & \multicolumn{1}{l|}{\textbf{StarCoder}} & \multicolumn{1}{l|}{\textbf{WizardCoder}} & \multicolumn{1}{l|}{\textbf{Mistral}} & \multicolumn{1}{l|}{\textbf{Phi-2}} & \multicolumn{1}{l|}{\textbf{DeepSeek-Coder}} & \multicolumn{1}{l|}{\textbf{CodeLlama}} & \multicolumn{1}{l|}{\textbf{StarCoder}} & \multicolumn{1}{l|}{\textbf{WizardCoder}} & \multicolumn{1}{l|}{\textbf{Mistral}} & \multicolumn{1}{l|}{\textbf{Phi-2}} \\
    \cmidrule{4-15}      &   &   & \multicolumn{6}{c|}{\textbf{Recall}} & \multicolumn{6}{c|}{\textbf{Accuracy}} \\
    \midrule
    CWE-119 & 1549 & 128 & 0.570 & \textbf{0.719} & 0.625 & 0.359 & 0.453 & 0.492 & 0.801 & 0.735 & 0.777 & \textbf{0.839} & 0.785 & 0.792 \\
    CWE-20 & 1082 & 80 & 0.609 & \textbf{0.844} & 0.609 & 0.313 & 0.297 & 0.578 & 0.804 & 0.735 & 0.758 & \textbf{0.866} & 0.793 & 0.814 \\
    CWE-264 & 800 & 64 & 0.763 & \textbf{0.900} & 0.613 & 0.400 & 0.525 & 0.700 & 0.839 & 0.795 & 0.759 & \textbf{0.850} & 0.810 & 0.846 \\
    CWE-399 & 697 & 35 & 0.815 & \textbf{0.926} & 0.630 & 0.481 & 0.556 & 0.704 & 0.885 & 0.821 & 0.815 & \textbf{0.901} & 0.854 & 0.877 \\
    CWE-125 & 582 & 29 & 0.586 & \textbf{0.931} & 0.621 & 0.414 & 0.517 & 0.517 & 0.808 & 0.746 & 0.771 & \textbf{0.830} & 0.763 & 0.765 \\
    CWE-200 & 573 & 27 & 0.514 & \textbf{0.743} & 0.600 & 0.286 & 0.343 & 0.457 & 0.829 & 0.743 & 0.770 & \textbf{0.841} & 0.784 & 0.812 \\
    CWE-189 & 442 & 21 & 0.625 & \textbf{0.813} & \textbf{0.813} & 0.563 & 0.500 & 0.750 & 0.853 & 0.776 & 0.828 & \textbf{0.891} & 0.833 & 0.869 \\
    CWE-362 & 413 & 16 & 0.200 & \textbf{0.800} & 0.600 & 0.200 & 0.200 & 0.200 & 0.847 & 0.794 & 0.821 & \textbf{0.908} & 0.821 & 0.855 \\
    CWE-416 & 406 & 12 & 0.667 & \textbf{0.750} & 0.667 & 0.333 & 0.500 & 0.500 & 0.835 & 0.796 & 0.773 & \textbf{0.884} & 0.828 & 0.820 \\
    CWE-476 & 367 & 11 & 0.571 & \textbf{1.000} & 0.429 & 0.286 & 0.286 & 0.143 & 0.782 & 0.687 & 0.706 & \textbf{0.820} & 0.719 & 0.725 \\
    \bottomrule
    \end{tabular}%
    }
  \label{tab:rq1_llm_top_10}%
\end{table*}%

\subsection{RQ-1: Evaluating Vulnerability Detection of LLMs}
In this RQ, we first investigate the vulnerability detection of LLMs and make a comparison with the existing state-of-the-art (SOTA) approaches. 
Then, we conduct a more detailed analysis of the results, comparing the detection performance of LLMs under the Top-10 CWE types.


\noindent
\textbf{\underline{Experimental Setting.}} 
We instruct LLMs with the following task description to tell it to act as a vulnerability detector.

\noindent
\intuition{
\textbf{Task Description:} If this C code snippet has vulnerabilities, output Yes; otherwise, output No.
}

\xy{In addition to pre-trained LMs, we also consider the following five SOTA baselines:
Devign~\cite{zhou2019devign}, {Reveal}~\cite{chakraborty2021deep}, {IVDetect}~\cite{li2021vulnerability}, LineVul~\cite{fu2022linevul}, and SVulD~\cite{ni2023distinguishing}.
These baselines can be divided into two groups: graph-based (i.e., Devign, Reveal, and  IVDetect) and transformer-based (i.e., pre-trained LMs, LineVul, and SVulD).} Besides, in order to comprehensively compare the performance among baselines and LLMs, we consider four widely used performance measures (i.e., Precision, Recall, F1-score, and Accuracy) and conduct experiments on the popular dataset.
Since graph-based approaches need to obtain the structure information (e.g., control flow graph (CFG), data flow graph (DFG)) of the studied functions, we adopt the same toolkit with \textit{Joern} to transform functions.
The functions are dropped out directly if they cannot be transformed by \textit{Joern} successfully.
Finally, the filtered dataset (shown in Table~\ref{tab:dataset}) is used for evaluation.
We follow the same strategy to build the training data, validation data, and testing data from the original dataset with previous work does~\cite{fu2022linevul, ni2022defect}. 
Specifically, 80\% of functions are treated as training data, 10\% of functions are treated as validation data, and the left 10\% of functions are treated as testing data. 
We also keep the distribution as same as the original ones in training, validation, and testing data.
\xy
{
We undersample the non-vulnerable functions to produce approximately balanced training data at the function level, while the validation and testing data remain in the original imbalanced ratio.
}Apart from presenting the overall performance comparison, we also give the detailed performance of LLMs on the Top-10 CWE types for a better analysis.

\noindent
\xy
{
\textbf{\underline{Results.}}
\textbf{[A] LLMs vs. Baselines.}
Table~\ref{tab:rq1_llm_vs_sota} shows the overall performance measures between LLMs and eleven baselines and the best performances are highlighted in bold.
According to the results in Table~\ref{tab:rq1_llm_vs_sota}, we can obtain the following observations:
}

\xy
{
\textbf{(1) Fine-tuned LLMs have poor performance compared with transformer-based approaches when considering \textit{F1-score}, \textit{Precision}, and \textit{Accuracy}.}
In particular, SVulD obtains 0.336, 0.282, and 0.915 in terms of F1-score, Precision, and Accuracy, which surpass the fine-tuned LLMs by 24.4\% to 57.0\%, 64.0\% to 108.9\%, and 6.3\% to 20.2\% in terms of F1-score, Precision, and Accuracy, respectively.
Notably, the F1-score performance of LineVul is significantly lower (0.272) than that reported in the original paper (0.910).
We further analyze this discrepancy in Section~\ref{sec:discussion_1}.
}

\xy
{
\textbf{(2) The performance of fine-tuned LLMs is comparable to graph-based approaches.}
For example, in terms of F1-score, fine-tuned LLMs achieve a range of 0.214 to 0.270.
In comparison, graph-based approaches achieve a range of 0.200 to 0.232. 
}

\xy
{
\textbf{(3) LLMs under few-shot setting have poor performance compared with baselines.}
LLMs ranging from 2.7B to 34B parameters perform less favorably than baselines in terms of F1-score and Precision.
However, as for Accuracy, SVulD (transformer-based) obtains the best performance (0.915) and DeepSeek-Coder 6.7B under few-shot setting achieves a performance of 0.823, which is better than the three graph-based approaches. 
}


\find{
    \xy
    {
    \textbf{Finding-1}. 
    LLMs can detect vulnerabilities, but fine-tuned LLMs perform weaker than transformer-based approaches. 
    Considering the computational resources and time costs of deploying LLMs, transformer-based approaches for vulnerability detection are a more efficient choice.
    }
}

\textbf{[B] Fine-Tuning vs. Few-Shot.} The experimental results are presented in Table~\ref{tab:rq1_llm_vs_sota}. 
Based on these experimental findings, we can draw the following observations: 
(1) LLMs fine-tuned for vulnerability detection demonstrate superior performance on the task compared to LLMs in the few-shot setting. 
The average F1-score and average Precision have doubled, while the average Recall has also shown improvement.
(2) LLMs with more parameters typically exhibit better performance.
For example, CodeLlama 34B improves upon CodeLlama 7B by 19.4\%, 34.5\%, and 37.0\% in terms of F1-score, Precision, and Accuracy, respectively.
However, different LLMs may exhibit performance variations due to differences in model design and the quality of pre-training data. 
(3) Phi-2 achieves performance approximating that of other LLMs with 7 billion parameters, even with a parameter size of 2.7 billion. 
This may be attributed to the higher quality of its pre-training data.

\find{
    \xy
    {
    \textbf{Finding-2}.
    After fine-tuning, the detection capability of LLMs has improved. Larger models usually perform better, but performance can also be influenced by model design and pre-training data.
    Therefore, fine-tuning the LLM on domain-specific data before using it as a vulnerability detector is necessary.
    }
}
\vspace{0.2cm}

\xy
{
\textbf{[C] The comparisons of Top-10 CWE types between LLMs.}
Table~\ref{tab:rq1_llm_top_10} shows the detailed comparisons of Top-10 CWE types between fine-tuned LLMs.
In this table, we highlight the best performance for each performance metric in bold.
According to the results, we can achieve the following observations:
(1) In most cases, CodeLlama obtains better performance than other LLMs in terms of F1-score, Precision, and Recall. 
Different LLMs have certain advantages in different CWE types, complementing each other.
(2) Considering the performance of F1-score, Precision, and Recall, CodeLlama achieves the best performances on CWE-125 (\textit{``Out-of-bounds Read''}), CWE-362 (\textit{``Concurrent Execution using Shared Resource with Improper Synchronization ('Race Condition')''}), and CWE-476 (\textit{``NULL Pointer Dereference''}), which indicates CodeLlama is exceptionally skilled at detecting and mitigating vulnerabilities related to memory handling and synchronization issues.
}

\find{
\xy
{
\textbf{Finding-3}. 
In general, different LLMs complementing each other, while CodeLlama obtains better performance in terms of F1-score, Precision, and Recall. 
}
}


\subsection{RQ-2: Evaluating Vulnerability Assessment of LLMs}

In this RQ, we delineate two task descriptions for vulnerability assessment: (1) code-based and (2) code-based with additional key information.
We compare the performance of LLMs in both task descriptions for vulnerability assessment and concurrently conduct a case study to illustrate the effectiveness of incorporating key important information.

\noindent
\textbf{\underline{Experimental Setting.}}
We instruct LLM with the following task descriptions (i.e., Task Description 1 and Task Description 2) to tell it to act as a vulnerability assessor.
We first provide LLM with the vulnerable codes to explore its performance (Task Description 1).
Moreover, we provide LLM with some key important information, including the CVE description, the project, the commit message as well as the file name when the vulnerable code exists to investigate the performance differences (Task Description 2).

\noindent
\intuition{
\textbf{Task Description 1:}
Provide a qualitative severity rating of CVSS v2.0 for the vulnerable C code snippet.

\textbf{Task Description 2:}
Provide a qualitative severity rating of CVSS v2.0 for the vulnerable C code snippet (with additional information).
}

\noindent
\xy
{
\textbf{\underline{Results.}} 
Table~\ref{tab:rq2_assessment} shows the detailed results of LLMs and six baselines on vulnerability assessment.
Based on these experimental results, we can observe a significant improvement in the vulnerability assessment capability of LLMs after fine-tuning.
Specifically, the accuracy has increased from 0.282 to 0.424, reaching a range of 0.759 to 0.860, while precision has improved from 0.296 to 0.355, now ranging from 0.512 to 0.854. 
This underscores the necessity of fine-tuning in vulnerability assessment task. 
Overall, fine-tuned code-related LLMs outperform pre-trained LMs in vulnerability assessment.
It is worth noting that DeepSeek-Coder, after fine-tuning, achieves the best performance compared to other LLMs and pre-trained LMs. 
If researchers need to perform tasks such as vulnerability assessment with LLM, 
fine-tuning DeepSeek-Coder is a more efficient choice.
We also find that Mistral exhibits a relatively smaller improvement after fine-tuning, which aligns with our expectations, as it is a general LLM. 
}

\begin{table}[htbp]
  \centering
  \caption{\xy{The comparison between LLMs and six baselines on software vulnerability assessment (RQ2)}}
  \resizebox{\linewidth}{!}
  {
    \begin{tabular}{lcccc}
    \toprule
    \textbf{Methods} & \textbf{F1-score} & \textbf{Recall} & \textbf{Precision} & \textbf{Accuracy} \\
    \midrule
    CodeBERT & 0.753 & 0.730 & 0.788 & 0.828 \\
    GraphCodeBERT & 0.701 & 0.666 & 0.772 & 0.802 \\
    UniXcoder & 0.745 & 0.761 & 0.734 & 0.817 \\
    PLBART & 0.735 & 0.741 & 0.731 & 0.789 \\
    CodeT5 & 0.743 & 0.750 & 0.741 & 0.817 \\
    CodeT5+ & 0.706 & 0.677 & 0.755 & 0.789 \\
    \midrule
    \rowcolor{lightgray} \textbf{Fine-Tuning Setting} &&&& \\
    DeepSeek-Coder 6.7B & \textbf{0.814} & \textbf{0.785} & \textbf{0.854} & \textbf{0.860} \\
    CodeLlama 7B & 0.768 & 0.749 & 0.794 & 0.827 \\
    StarCoder 7B & 0.671 & 0.677 & 0.666 & 0.764 \\
    WizardCoder 7B & 0.793 & 0.778 & 0.813 & 0.842 \\
    Mistral & 0.525 & 0.539 & 0.512 & 0.759 \\
    Phi-2 & 0.747 & 0.732 & 0.767 & 0.802 \\
    \midrule
    \rowcolor{lightgray} \textbf{Few-Shot Setting} &&&& \\
    DeepSeek-Coder 6.7B & 0.229 & 0.339 & 0.310 & 0.262 \\
    DeepSeek-Coder 33B & 0.290 & 0.323 & 0.336 & 0.335 \\
    CodeLlama 7B & 0.310 & 0.331 & 0.334 & 0.373 \\
    CodeLlama 34B & 0.265 & 0.323 & 0.327 & 0.294 \\
    StarCoder 7B & 0.265 & 0.342 & 0.333 & 0.330 \\
    StarCoder 15.5B & 0.285 & 0.315 & 0.329 & 0.326 \\
    WizardCoder 7B & 0.244 & 0.351 & 0.336 & 0.250 \\
    WizardCoder 34B & 0.306 & 0.330 & 0.325 & 0.379 \\
    Mistral & 0.283 & 0.308 & 0.296 & 0.424 \\
    Phi-2 & 0.269 & 0.359 & 0.355 & 0.282 \\
    \bottomrule
    \end{tabular}%
    }
  \label{tab:rq2_assessment}%
\end{table}%

\begin{table*}[htbp]
  \centering
  \caption{A vulnerable code for CodeLlama to assess with different prompts (RQ2)}
  \resizebox{\linewidth}{!}
  {
    \begin{tabular}{|l|p{48em}|}
    \hline
    \multicolumn{2}{|p{55em}|}{\textbf{Improper Restriction of Operations within the Bounds of a Memory Buffer Vulnerability (CWE-119) in Linux}} \\
    \hline
    \textbf{Task Description 1} & {{Provide a qualitative severity ratings of CVSS v2.0 for the vulnerable C code snippet.}} \\
    \hline
    \textbf{Input 1} & An example of a C code snippet with vulnerabilities (CVE-2011-2517). 
    \\
    \hline
    \textbf{Response 1} &  \textbf{Severity: Medium} \\
    \hline
    \textbf{Task Description 2} & {Provide a qualitative severity rating of CVSS v2.0 for the vulnerable C code snippet (with additional information).} \\
    \hline
    \textbf{Input 2} & \textbf{Project}: Linux
    \newline{}\textbf{File Name}: net/wireless/nl80211.c
    \newline{}\textbf{CVE Description}: Multiple buffer overflows in net/wireless/nl80211.c in the Linux kernel before 2.6.39.2 allow local users to gain privileges by leveraging the CAP\_NET\_ADMIN capability during scan operations with a long SSID value.
    \newline{}\textbf{Commit Message}:
    nl80211: fix check for valid SSID size in scan operations.
    In both trigger\_scan and sched\_scan operations, we were checking for the SSID length before assigning the value correctly. Since the memory was just kzalloc'ed, the check was always failing and SSID with
    over 32 characters were allowed to go through.
    This was causing a buffer overflow when copying the actual SSID to the proper place.
    This bug has been there since 2.6.29-rc4. \\
    \hline
    \textbf{Response 2} & \textbf{Severity: High} \\
        \hline
    \textbf{Analysis } & The true Severity is High. After providing additional key information, CodeLlama output for the Severity changed from Medium to High. \\
        \hline
    \end{tabular}
    }
  \label{tab:rq2_example}%
\end{table*}%

\find{
\xy
{
\textbf{Finding-4.} 
Overall, fine-tuned code-related LLMs outperform pre-trained LMs in vulnerability assessment.
When resources permit, fine-tuning DeepSeek-Coder 6.7B for vulnerability assessment is optimal, as it outperforms the pre-trained LMs across four metrics.
}
}
\vspace{0.2cm}



\textbf{Case Study.} To illustrate the effectiveness of key important information, we present an instance of a vulnerability (CWE-119) in Big-Vul that is exclusively assess by CodeLlama, as depicted in Table~\ref{tab:rq2_example}. 
This example is a vulnerability in the Linux project, categorized under CWE-119 (Improper Restriction of Operations within the Bounds of a Memory Buffer Vulnerability).
In an initial assessment without critical information, CodeLlama did not fully grasp the severity of this vulnerability and labeled it as ``Medium''.
However, with the provision of crucial details, CodeLlama can more accurately evaluate the risk level of this vulnerability.
The CVE description for this vulnerability highlights multiple buffer overflows in the net/wireless/nl80211.c file of the Linux kernel prior to version 2.6.39.2. 
These vulnerabilities allow local users to gain elevated privileges by leveraging the CAP NET ADMIN capability during scan operations with an excessively long SSID value. 
In this scenario, the lack of proper validation of the SSID length leads to buffer overflows, enabling attackers to exploit the vulnerability, escalate privileges, and execute malicious code. 
The commit message described that this bug has existed since version 2.6.29-rc4 of the Linux kernel.
Given this information, CodeLlama reassesses the risk level of this vulnerability as ``High''. 
This is because it allows attackers to escalate privileges and execute malicious code, and it has persisted for a considerable period of time. 
It is crucial to address and patch this vulnerability promptly by updating the operating system or kernel to ensure security.

\begin{figure*}[!htbp]
    \centering
    \includegraphics[width=\linewidth]{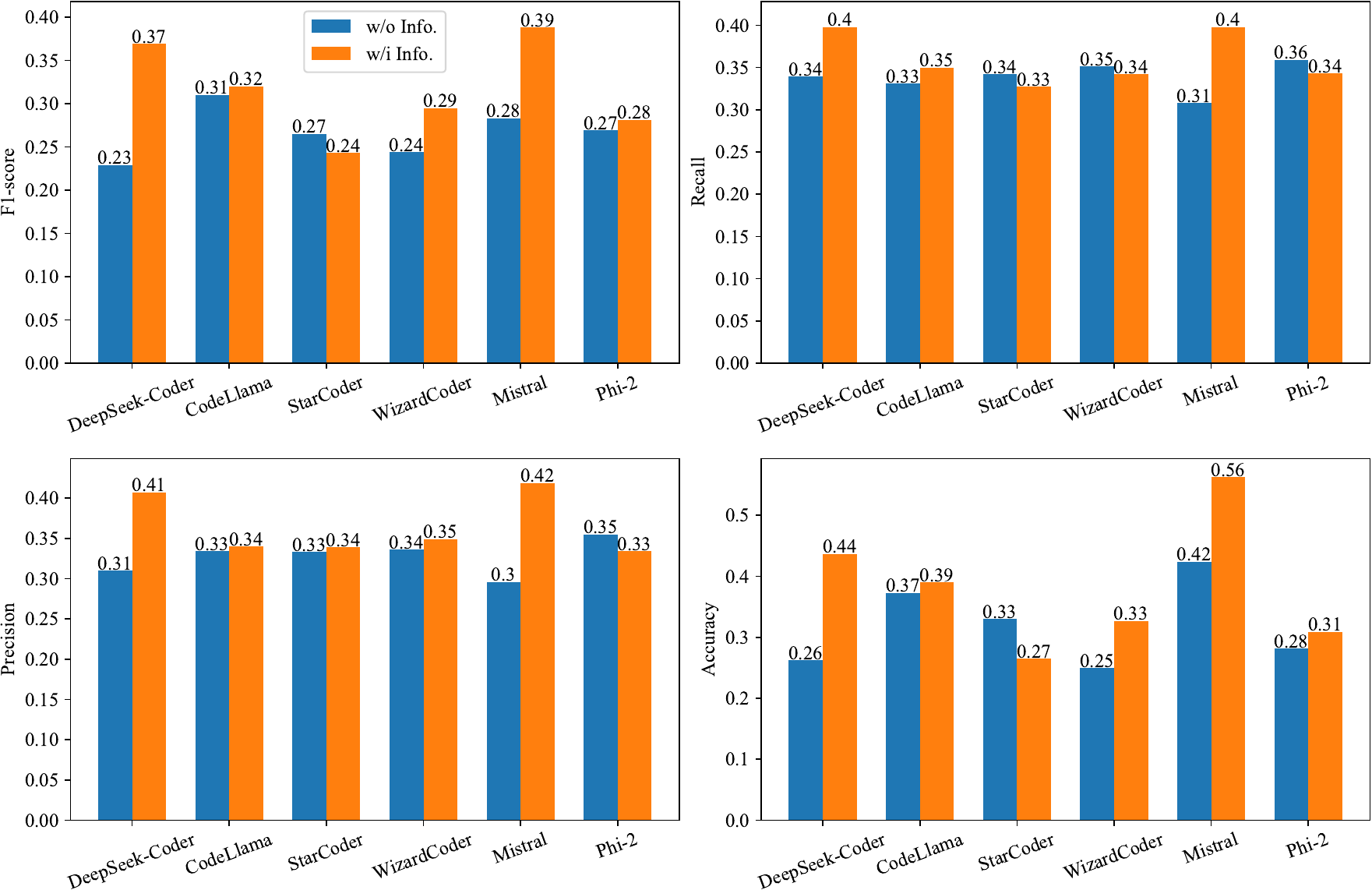}
    \caption{\xy{The impact of key important information on LLM Vulnerability Assessment (RQ2)}}
    \label{fig:rq2_chart}
\end{figure*}

\xy
{
To compare the vulnerability assessment capabilities of LLMs after providing key information, we have created a performance comparison bar chart, as shown in Fig.~\ref{fig:rq2_chart}. 
LLMs have limited capacity for assessing vulnerability severity based solely on source code. 
However, when provided with key important information, most LLMs (i.e., DeepSeek-Coder, CodeLlama, WizardCoder, and Mistral) exhibit significantly improved vulnerability assessment capabilities, particularly in terms of the Accuracy metric.
The Accuracy has increased from the range of 0.26-0.42 to the range of 0.27-0.56.
StarCoder and Phi-2 are showing a declining trend, and we believe this may be attributed to the addition of key information, resulting in an increase in the number of input tokens. 
These LLMs may not excel in handling excessively long text sequences, and we analyze this further in Section~\ref{sec:discussion_2}. 
In contrast, DeepSeek-Coder and Mistral exhibit significant improvements, possibly due to their proficiency in handling long sequential text.
}

\find{
\textbf{Finding-5.} LLMs have the capacity for assessment of vulnerability severity based on source code, and can be improved by providing more context information.
}

\subsection{RQ-3: Evaluating Vulnerability Location of LLMs}
\label{rq3}

In this RQ, we first outline how to assess the vulnerability location capabilities of LLMs. 
Then, we proceed to compare the vulnerability location abilities of LLMs across different settings, both at a general level and in detail, and analyze the reasons behind the observed differences.

\noindent
\textbf{\underline{Experimental Setting.}} 
We select the vulnerable functions with information on vulnerable lines from the testing set for the evaluation and instruct LLM with the following task description to explore its vulnerability location performance.

\noindent
\intuition{
\textbf{Task Description:} Provide a vulnerability location result for the vulnerable C code snippet.
}

\xy
{
For the fine-tuning setting of LLMs and pre-trained LMs, we treat the vulnerability location task as a binary classification problem, determining whether each line of code is vulnerable or not.
For the few-shot setting, a specific vulnerable function may contain one or several vulnerable lines, and LLM may also predict one or several potential vulnerable lines ($Lines_{predict}$). 
We convert $Lines_{predict}$ into a binary classification format.
For example, if a given vulnerable function consists of five lines and contains two vulnerable lines \textbf{[2, 3]}, and the LLM predicts one potential vulnerable line \textbf{[2]}, we convert this to a binary classification format as \textbf{[0, 0, 1, 0, 0]} for ease of computation.
To better evaluate the vulnerability location performance of LLM on a specific vulnerable function, we consider five widely used performance measures (i.e., Precision, Recall, F1-score, Accuracy, and FPR).
}

\xy
{
In addition to pre-trained LMs, we also consider the following four SOTA baselines:
Devign~\cite{zhou2019devign}, {Reveal}~\cite{chakraborty2021deep}, {IVDetect}~\cite{li2021vulnerability}, and LineVul~\cite{fu2022linevul}.
For the graph-based approaches (i.e., Devign, Reveal, and IVDetect), we use GNNExplainer~\cite{ying2019gnnexplainer,hu2023interpreters} for vulnerability location. 
We compare the performance of LLMs and these baselines using Top-k Accuracy, as employed in previous works~\cite{fu2022linevul,hu2023interpreters}.
}

\begin{table}[htbp]
\centering
\caption{\xy{The comparison between LLMs and six baselines on software vulnerability location (RQ3)}}
\resizebox{\linewidth}{!}
{
\begin{tabular}{lccccc}
    \toprule
    \textbf{Methods} & \textbf{F1-score} & \textbf{Recall} & \textbf{Precision} & \textbf{Accuracy} & \textbf{FPR} \\
    \midrule
    CodeBERT & 0.470 & \textbf{0.514} & 0.433 & 0.879 & 0.078 \\
    GraphCodeBERT & 0.483 & 0.477 & 0.489 & 0.893 & 0.058 \\
    UniXcoder & 0.460 & 0.384 & 0.575 & 0.908 & 0.032 \\
    PLBART & 0.436 & 0.416 & 0.458 & 0.886 & 0.058 \\
    CodeT5 & 0.493 & 0.408 & 0.623 & 0.914 & 0.028 \\
    CodeT5+ & 0.303 & 0.207 & 0.565 & 0.902 & \textbf{0.018} \\
    \midrule
    
    \rowcolor{lightgray}\textbf{Fine-Tuning Setting} &  &  &  &  &  \\
    DeepSeek-Coder 6.7B & 0.437 & 0.332 & 0.640 & 0.912 & 0.021 \\
    CodeLlama 7B & 0.504 & 0.396 & \textbf{0.691} & \textbf{0.919} & 0.021 \\
    StarCoder 7B & 0.245 & 0.169 & 0.443 & 0.893 & 0.024 \\
    WizardCoder 7B & \textbf{0.520} & 0.427 & 0.664 & 0.918 & 0.025 \\
    Mistral & 0.314 & 0.384 & 0.266 & 0.827 & 0.122 \\
    Phi-2 & 0.458 & 0.361 & 0.629 & 0.912 & 0.025 \\
    \midrule
    
    \rowcolor{lightgray}\textbf{Few-Shot Setting} &  &  &  &  &  \\
    DeepSeek-Coder 6.7B & 0.111 & 0.111 & 0.112 & 0.852 & 0.081 \\
    DeepSeek-Coder 33B & 0.110 & 0.112 & 0.108 & 0.849 & 0.084 \\
    CodeLlama 7B & 0.082 & 0.063 & 0.116 & 0.882 & 0.043 \\
    CodeLlama 34B & 0.115 & 0.090 & 0.158 & 0.884 & 0.044 \\
    StarCoder 7B & 0.088 & 0.066 & 0.134 & 0.887 & 0.039 \\
    StarCoder 15.5B & 0.095 & 0.078 & 0.120 & 0.876 & 0.052 \\
    WizardCoder 7B & 0.082 & 0.063 & 0.120 & 0.884 & 0.042 \\
    WizardCoder 34B & 0.096 & 0.072 & 0.145 & 0.887 & 0.039 \\
    Mistral & 0.086 & 0.065 & 0.127 & 0.885 & 0.040 \\
    Phi-2 & 0.073 & 0.053 & 0.116 & 0.885 & 0.037 \\
    \bottomrule
\end{tabular}
\label{tab:rq_3_location}
}
\end{table}

\noindent
\xy
{
\textbf{\underline{Results.}}
Table~\ref{tab:rq_3_location} presents the overall performance of vulnerability location between LLMs and seven baselines.
Based on this table, we can achieve the following observations: 
\textbf{(1) Fine-tuning can greatly enhance the vulnerability location capabilities of LLMs.}
For example, after fine-tuning, CodeLlama 7B's F1-score increases from 0.082 to 0.504, recall increases from 0.063 to 0.396, precision increases from 0.116 to 0.691, accuracy increases from 0.882 to 0.919, and FPR decreases from 0.043 to 0.021.
\textbf{(2) Code-related LLMs often outperform pre-trained LMs in terms of F1-score, precision, and FPR.}
For example, 
CodeLlama 7B outperforms the pre-trained LMs, which are averaged over six different models, by 14.3\%, 31.9\%, and 54.3\% in terms of F1-score, precision, and FPR, respectively.
}

\begin{figure}[!htbp]
    \centering
    \includegraphics[width=\linewidth]{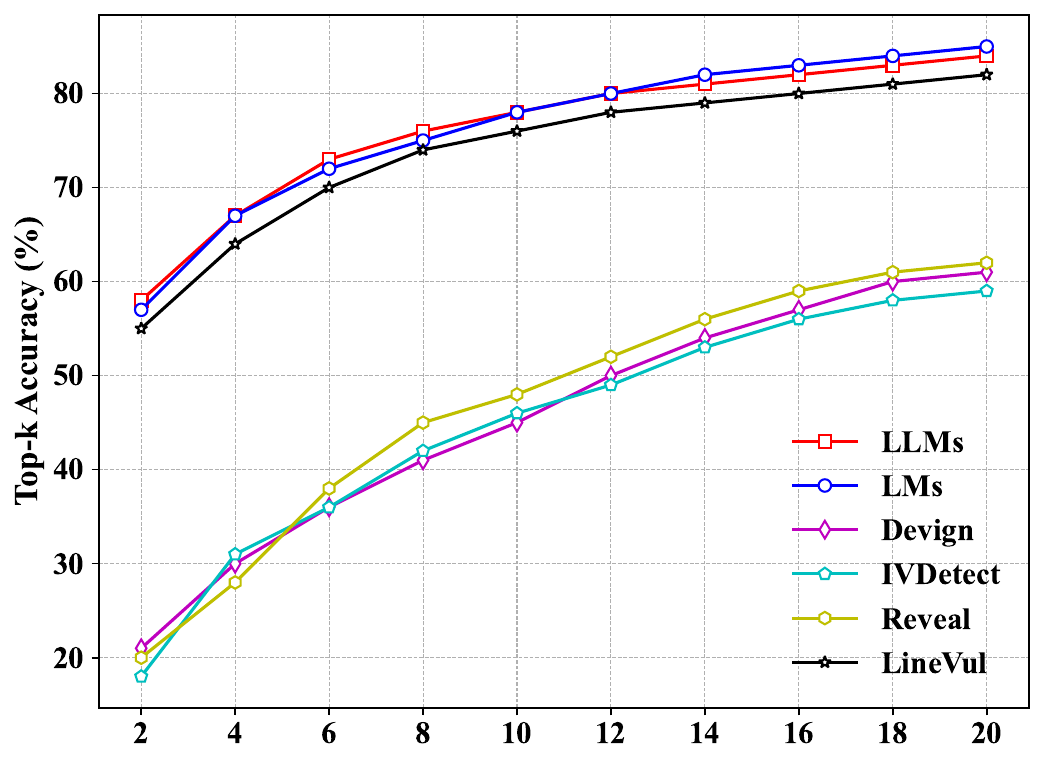}
    \caption{\xy{The interpretation results for different vulnerability locators with k from 2 to 20 (RQ3)}}
    \label{fig:rq3}
\end{figure}

\xy
{
The Top-k Accuracy results of the interpreters are shown in Fig.~\ref{fig:rq3}, where the x-axis represents k and the y-axis represents Top-k Accuracy (\%). 
For comparison, we average the results of all LLMs and pre-trained LMs. 
We can observe that the performance of these graph-based methods does not show significant differences but is considerably weaker than that of pre-trained LMs, LLMs, and LineVul. 
Although pre-trained LMs achieve the highest accuracy at k=20, the difference between them and LLMs is not substantial.
}

\begin{table*}[htbp]
  \centering
  \caption{\xy{An example of generating vulnerability description for vulnerable code (RQ4)}}
  \resizebox{\linewidth}{!}
  {
    \begin{tabular}{|l|p{40.165em}|}
    \hline
    \multicolumn{2}{|l|}{\textbf{Describe vulnerability code (CWE-119, Improper Restriction of Operations within the Bounds of a Memory Buffer)}} \\
    \hline
    \textbf{Task Description} & {Provide a CVE description for the vulnerable C code snippet.} \\
    \hline
    \textbf{Source Code} & {An example of a C code snippet with vulnerabilities. \textbf{CVE ID: CVE-2016-0841}.} \\
    \hline
    \textbf{Response From CodeT5+} & 
    libmedia 
    \textcolor{orange}{\textbf{in mediaserver in Android 4.x before 4.4.4, 5.0.x before 5.0.2, 5.1.x before 5.1.1, and 6.x before}} 
    2016-08-01 has certain incorrect declarations, 
    \textcolor{orange}{\textbf{which allows remote attackers to execute arbitrary code or cause a denial of service}} (NULL pointer dereference or \textcolor{orange}{\textbf{memory corruption) via a crafted media file, aka internal bug}}
    28166152. \\
    \hline
    \textbf{Response From CodeLlama} & \textcolor{orange}{\textbf{media/libmedia/MediaMetadataRetriever.cpp in mediaserver in Android 4.x before 4.4.4, 5.0.x before 5.0.2, 5.1.x before 5.1.1, and 6.x before 2016-04-01}} does not validate the data size, 
    \textcolor{orange}{\textbf{which allows attackers to execute arbitrary code or cause a denial of service (memory corruption) via a crafted media file, aka internal bug}} 28799341.
    Specifically, the constructor calls the native function createMetadataRetriever in libstagefright, and later calls the getFrameAtTime function, aka internal bug 29244311. The getFrameAtTime call triggers a memory corruption warning, which allows an attacker to execute arbitrary code via a crafted video file, aka internal ... \\
    \hline
    \textbf{CVE Description} & media/libmedia/mediametadataretriever.cpp in mediaserver in Android 4.x before 4.4.4, 5.0.x before 5.0.2, 5.1.x before 5.1.1, and 6.x before 2016-04-01 mishandles cleared service binders, which allows remote attackers to execute arbitrary code or cause a denial of service (memory corruption) via a crafted media file, aka internal bug 26040840. \\
    \hline
    \end{tabular}
    }
  \label{tab:rq4_example}%
\end{table*}%

\begin{figure}[!htp]
    \centering
    \includegraphics[width=\linewidth]{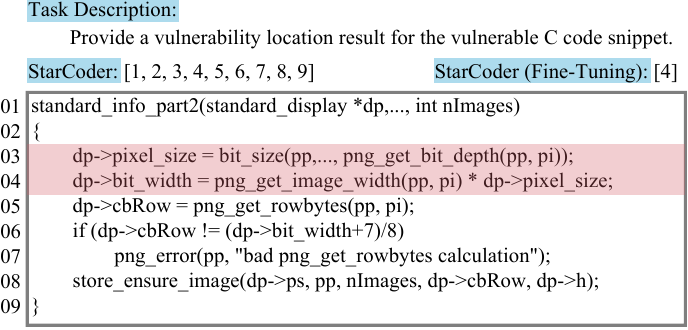}
    \caption{An example to demonstrate the limitations of StarCoder in vulnerability location (RQ3)}
    \label{fig:rq3_example}
\end{figure}

\xy
{
\textbf{Case Study.} 
We find that LLMs under few-shot setting have limitations, and in some exceptional cases, they tend to output more vulnerable lines, even if these lines do not contain vulnerabilities. 
We take StarCoder as an example, Fig.~\ref{fig:rq3_example} depicts a vulnerability code snippet from the Big-Vul dataset, with the vulnerability behavior occurring in lines 3 and 4.
However, StarCoder tends to output more vulnerability lines, such as \textbf{``[1, 2, 3, 4, 5, 6, 7, 8, 9]''}, whereas after fine-tuning, StarCoder becomes more cautious and only predicts \textbf{``[4]''}.
Note that we convert the model's predictions into a specific format, i.e., transform \textbf{``[0, 0, 0, 1, 0, 0, 0, 0, 0]''} to \textbf{``[4]''}.
}

\find{
\xy
{
\textbf{Finding-6.}
Few-shot setting exposes LLM's limitations, and fine-tuning can greatly enhance the vulnerability location capabilities of LLMs.
}

\xy
{
\textbf{Finding-7.} 
Fine-tuning code-related LLMs as vulnerability locators is beneficial, as they can outperform pre-trained LMs in terms of F1-score, precision, and FPR.
}
}

\subsection{RQ-4: Evaluating Vulnerability Description of LLMs}

In this RQ, we employ the ROUGH metric to evaluate the LLMs' vulnerability description capabilities. 
We conduct a detailed statistical analysis of LLMs' abilities and also perform a case study to provide a comprehensive assessment of their performance in describing vulnerabilities.

\noindent
\textbf{\underline{Experimental Setting.}}
We instruct LLMs with a designated task description, guiding them to perform the role of a vulnerability descriptor. Table~\ref{tab:rq4_example} illustrates an example of our approach to evaluating LLMs' proficiency in conducting vulnerability descriptions.

\noindent
\intuition{
{\textbf{Task Description}}: 
Provide a CVE description for the vulnerable C code snippet.
}

To evaluate the precision of the generated CVE description, we adopt the widely used performance metric ROUGE~\cite{lin2004rouge}, which is a set of metrics and is used for evaluating automatic summarization and machine translation software in natural language processing. 
The metrics compare an automatically produced summary or translation against a reference or a set of references (human-produced) summary or translation.
Here, we totally consider three settings: \textit{1}, \textit{2}, and \textit{L}.

\noindent
\xy
{
\textbf{\underline{Results.}}
Table~\ref{tab:rq_4_description} represents the vulnerability description capabilities of LLMs and six baselines. 
According to the results, we can obtain the following observations:
\textbf{(1) LLMs exhibit significantly weaker performance in generating vulnerability descriptions compared to pre-trained LMs.} 
For instance, pre-trained LMs achieve an average performance of 0.600, 0.487, and 0.591 on ROUGE-1, ROUGE-2, and ROUGE-L, respectively, whereas fine-tuned LLMs only achieve an average of 0.406, 0.301, and 0.400 on the same metrics.
\textbf{(2) Fine-tuning can significantly enhance the performance of LLMs in vulnerability descriptions.}
After fine-tuning, there is a several-fold improvement in ROUGE-1, ROUGE-2, and ROUGE-L. 
This suggests that these LLMs possess strong learning capabilities and can extract more gains from historical data. 
(3) The low ROUGE-2 scores indicate that Phi-2 has a limited ability to generate accurate and relevant high-order n-grams (pairs of consecutive words) in vulnerability descriptions, indicating potential issues in capturing specific and detailed information.
}

\begin{table}[htbp]
  \centering
  \caption{\xy{The comparison of LLMs on software vulnerability description (RQ4)}}
  \resizebox{\linewidth}{!}
  {
    \begin{tabular}{lccc}
    \toprule
    \textbf{Methods} & \textbf{ROUGE-1} & \textbf{ROUGE-2} & \textbf{ROUGE-L} \\
    \midrule
    CodeBERT & 0.511 & 0.376 & 0.501 \\
    GraphCodeBERT & 0.538 & 0.406 & 0.528 \\
    UniXcoder & 0.658 & 0.558 & 0.650 \\
    PLBART & 0.447 & 0.313 & 0.437 \\
    CodeT5 & 0.700 & 0.604 & 0.693 \\
    CodeT5+ & \textbf{0.747} & \textbf{0.668} & \textbf{0.740} \\
    \midrule
    \rowcolor{lightgray} \textbf{Fine-Tuning Setting} &&&\\
    DeepSeek-Coder 6.7B & 0.434 & 0.325 & 0.425 \\
    CodeLlama 7B & 0.392 & 0.292 & 0.387 \\
    StarCoder 7B & 0.420 & 0.321 & 0.416 \\
    WizardCoder 7B & 0.425 & 0.327 & 0.419 \\
    Mistral & 0.453 & 0.347 & 0.448 \\
    Phi-2 & 0.313 & 0.196 & 0.305 \\
    \midrule
   \rowcolor{lightgray} \textbf{Few-Shot Setting} &&& \\
    DeepSeek-Coder 6.7B & 0.230 & 0.073 & 0.215 \\
    DeepSeek-Coder 33B & 0.219 & 0.066 & 0.203 \\
    CodeLlama 7B & 0.221 & 0.070 & 0.205 \\
    CodeLlama 34B & 0.258 & 0.094 & 0.242 \\
    StarCoder 7B & 0.243 & 0.084 & 0.229 \\
    StarCoder 15.5B & 0.255 & 0.089 & 0.241 \\
    WizardCoder 7B & 0.230 & 0.066 & 0.211 \\
    WizardCoder 34B & 0.276 & 0.111 & 0.261 \\
    Mistral & 0.290 & 0.095 & 0.267 \\
    Phi-2 & 0.210 & 0.056 & 0.194 \\
    \bottomrule
    \end{tabular}%
  \label{tab:rq_4_description}%
  }
\end{table}%

\xy
{
\textbf{Case Study.}
To demonstrate the capability of pre-trained LMs and LLMs in generating vulnerability descriptions, we present an example of a vulnerability (CWE-119) described by CodeT5+ and CodeLlama, as shown in Table~\ref{tab:rq4_example}.
This example represents a vulnerability within the Linux project, categorized as CWE-119 (Improper Restriction of Operations within the Bounds of a Memory Buffer Vulnerability). 
It is noteworthy that even when provided with only the code of the vulnerability, CodeT5+ produces text highly similar to the CVE description (highlighted in orange), indicating that pre-trained LMs are capable of comprehending the essence and crucial features of vulnerabilities and expressing this information in natural language.
}

\xy
{
Additionally, we find that CodeLlama's response is very similar to the CVE description, but with many additional details. 
We hypothesize that the poor performance of LLMs is not due to their inability to generate appropriate vulnerability descriptions, but rather because they tend to output tokens endlessly, even when they should stop. 
In contrast, pre-trained LMs typically stop at the appropriate points.
}

\xy
{
To further analyze this, we investigate the vulnerability description capabilities of LLMs after mitigating this issue. 
Using CodeLlama as an example, we randomly select 100 examples from the testing set and manually determine where the descriptions should terminate, trimming CodeLlama's output accordingly. 
We then calculate the ROUGE metrics for the trimmed outputs and compare them with the original results and those of CodeT5+.
The final results are presented in Table~\ref{tab:rq_4_discussion}, we find that after trimming, the ROUGE-1, ROUGE-2, and ROUGE-L scores for CodeLlama significantly improved, even nearing those of CodeT5+. 
This confirms our hypothesis that LLMs actually possess strong vulnerability description capabilities, but their performance is hindered by the tendency to output excessively.
}

\begin{table}[htbp]
  \centering
  \caption{\xy{The comparison of CodeT5+, CodeLlama, and CodeLlama-Trim on selected examples (RQ4)}}
  \resizebox{.95\linewidth}{!}
  {
    \begin{tabular}{lccc}
    \toprule
    \textbf{Methods} & \textbf{ROUGE-1} & \textbf{ROUGE-2} & \textbf{ROUGE-L} \\
    \midrule
    CodeT5+ & 0.730 & 0.644 & 0.722 \\
    CodeLlama & 0.366 & 0.266 & 0.360 \\
    CodeLlama-Trim & 0.625 & 0.523 & 0.616 \\
    \bottomrule
    \end{tabular}%
  \label{tab:rq_4_discussion}%
  }
\end{table}%

\find{
\xy
{
\textbf{Finding-8.}
LLMs exhibit significantly weaker performance in generating vulnerability descriptions compared to pre-trained LMs.
Therefore, fine-tuning pre-trained LMs for vulnerability detection is recommended.
}
}

\section{\xy{Discussion}}
\label{sec:discussion}

\xy
{
This section discusses open questions regarding the performance differences observed, the impact of input sequence length, and potential threats to the validity of our results.
}

\subsection{\xy{Analysis of Performance Difference}}
\label{sec:discussion_1}
\xy
{
In RQ1, for LineVul, there is a huge difference between the results obtained in this paper (i.e., 0.272 of F1-score) and the ones reported in original work (i.e., 0.910 of F1-score). 
To ensure a fair comparison, we first check the correctness of our LineVul reproduction by re-conducting the corresponding experiments using the original dataset provided by LineVul's official source and we obtain similar results. 
Then, we inspect each step of the data preprocessing process, as outlined in Section~\ref{lab:dataset}. 
In particular, this process involves three pre-processing in total: removing blank lines, removing comments, and trimming leading and trailing spaces from lines. 
We pre-process the original dataset of LineVul, re-train, and test the model under the same parameter settings. The results are shown in Table~\ref{tab:disussion_1} and we obtain the following conclusions: 
}




\begin{itemize}[leftmargin=*]
\item \xy{Our reproduced LineVul performs closely to the original one.} 
\item \xy{Removing blank lines and comments does not significantly affect LineVul's results.}
\item \xy{Trimming leading and trailing spaces from lines causes a drastic decrease in LineVul's performance.} 
\end{itemize}

\xy
{
Generally, for C/C++ source code, we know that removing leading and trailing spaces does not affect the code's semantics.
Thus, to verify whether it is general to other transformer-based models, we conduct another experiment on UniXcoder (another famous and widely used transformer-based pre-trained model) by adopting the same filtering operations. 
The results are presented in the right part of Table~\ref{tab:disussion_1}.
Table~\ref{tab:disussion_1} shows that the UniXcoder's performance closely resembled LineVul's before the third step of processing. 
However, after pre-processing, UniXcoder's performance similarly plummeted. Thus, we believe that such types of operation will have side impacts on transformer-based models since these methods pay attention to each token, though these tokens have no semantic meaning in the context of source code.
Based on this observation, we believe that the vulnerability detection effectiveness of LineVul after space removal is correct and the performance results are reasonable.
}

\begin{table*}[htbp]
\centering
\caption{\xy{The reproduced results for LineVul and UniXcoder}}
\resizebox{.9\linewidth}{!}{
    \begin{tabular}{l|cccc|cccc}
    \toprule
    \multirow{2.5}{*}{\textbf{Datasets}} & \multicolumn{4}{c|}{\textbf{LineVul}} & \multicolumn{4}{c}{\textbf{UniXcoder}} \\
    \cmidrule{2-9}
    & \textbf{F1-score} & \textbf{Accuracy} & \textbf{Recall} & \textbf{Precision} & \textbf{F1-score} & \textbf{Accuracy} & \textbf{Recall} & \textbf{Precision} \\
    \midrule
    {Original dataset} & 0.90 & 0.95 & 0.86 & 0.95 & 0.86 & 0.98 & 0.82 & 0.90 \\
    \midrule
    Remove empty lines & 0.85 & 0.98 & 0.79 & 0.93 & 0.85 & 0.98 & 0.82 & 0.88 \\
    Remove comments & 0.86 & 0.99 & 0.81 & 0.93 & 0.85 & 0.98 & 0.81 & 0.90 \\
    \rowcolor{lightgray} \textbf{Remove spaces} & \textbf{0.40} & \textbf{0.94} & \textbf{0.37} & \textbf{0.45} & \textbf{0.26} & \textbf{0.95} & \textbf{0.16} & \textbf{0.68} \\
    \bottomrule
    \end{tabular}
    \label{tab:disussion_1}
}
\end{table*}

\subsection{\xy{Analysis of Input Sequence Length}}
\label{sec:discussion_2}
\begin{figure}[!htbp]
    \centering
    \vspace{0.2cm}
    \includegraphics[width=\linewidth]{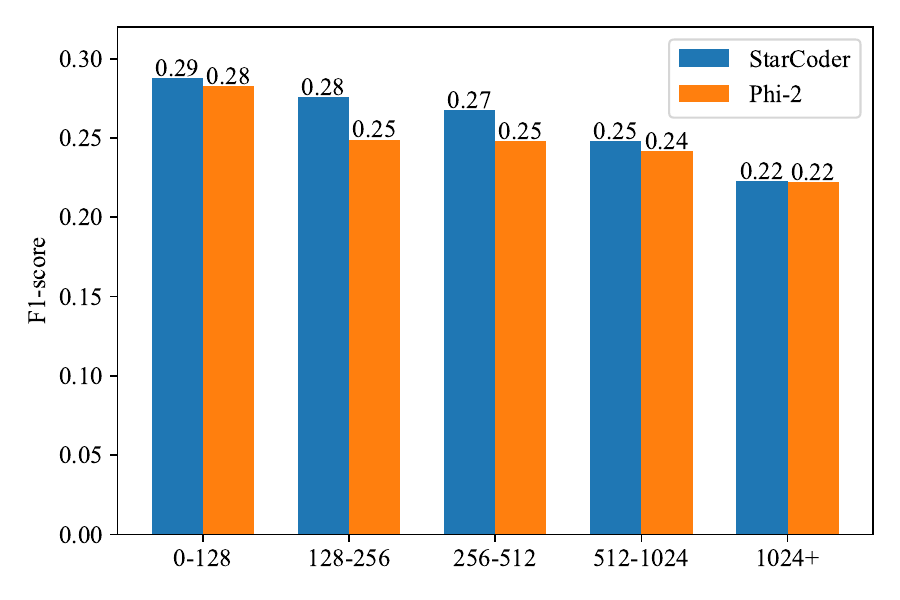}
    \caption{\xy{The variation of F1-scores for vulnerability assessment with respect to input sequence length}}
    \label{fig:discussion}
    \vspace{-0.2cm}
\end{figure}

\xy
{
In RQ2, we find that after adding key information, the performance of StarCoder and Phi-2 in vulnerability assessment actually weakened. 
We hypothesize that these LLMs may not excel in handling excessively long text sequences. 
Therefore, adding key information, which results in an increase in the number of input tokens, leads to a decline in performance.
In this section, we aim to analyze the performance of StarCoder and Phi-2 with respect to input sequence length to determine whether there is a performance decline as the input length increases.
As shown in Fig.~\ref{fig:discussion}, the horizontal axis represents the token length of the input sequence, and the vertical axis represents the F1-score of vulnerability assessment. 
We categorize the input token lengths into 0-128, 128-256, 256-512, 512-1024, and 1024+ (e.g., an input token length of 64 falls into the 0-128 range), and evaluate the vulnerability assessment performance of StarCoder and Phi-2 for each category. According to Fig.~\ref{fig:discussion}, we observe that as the input length increases, the F1-scores of both LLMs gradually decrease, revealing their significant limitations in assessing long sequences of vulnerable code. 
Therefore, in practical applications requiring the assessment of long sequences of vulnerable code, we may need to consider alternative optimization strategies or model choices to ensure accuracy and reliability.
}

\subsection{Threats to Validity}

\noindent
\textbf{Threats to Internal Validity} mainly contains in two-folds.
The first one is the design of a prompt to instruct LLMs to give out responses.
We design our prompt according to the practical advice~\cite{shieh2023best} which has been verified by many users online and can obtain a good response from LLMs.
Furthermore, LLMs will generate responses with some randomness even given the same prompt.
Therefore, we set ``temperature'' to 0, which will reduce the randomness at most and we try our best to collect all these results in two days to avoid the model being upgraded. 
The second one is about the potential mistakes in the implementation of studied baselines. 
To minimize such threats, we directly use the original source code shared by corresponding authors.

\noindent
\textbf{Threats to External Validity} may correspond to the generalization of the studied dataset.
To mitigate this threat, we adopt the most large-scale vulnerability dataset with diverse information about the vulnerabilities, which are collected from practical projects, and these vulnerabilities are recorded in the Common Vulnerabilities and Exposures (CVE).
However, we do not consider these vulnerabilities found recently.
Besides, we do not adopt another large-scale vulnerability dataset named SARD since it is built manually and cannot satisfy the distinct characteristics of the real world~\cite{hin2022linevd,chakraborty2021deep}.

\noindent
\textbf{Threats to Construct Validity} mainly correspond to the performance metrics in our evaluations.
To minimize such threats, we consider a few widely used performance metrics to evaluate the performance of LLMs on different types of tasks, e.g., Recall, Precision, and ROUGE.

\section{Conclusion}

This paper aims to comprehensively investigate the capabilities of LLMs for software vulnerability tasks as well as its impacts.
To achieve that, we adopt a large-scale vulnerability dataset (named Big-Vul) and then conduct several experiments focusing on four dimensions: 
\textbf{(1) Vulnerability Detection}, \textbf{(2) Vulnerability Assessment}, \textbf{(3) Vulnerability Location}, and \textbf{(4) Vulnerability Description}.
Overall, although LLMs show some ability in certain areas, they still need further improvement to be competent in software vulnerability-related tasks.
Our research conducts a comprehensive survey of LLMs' capabilities and provides a reference for enhancing its understanding of software vulnerabilities in the future.

\section*{Acknowledgements}
This work was supported by the National Natural Science Foundation of China (Grant No.62202419 and No. 62172214), the Ningbo Natural Science Foundation (No. 2022J184), and the Key Research and Development Program of Zhejiang Province (No.2021C01105).

\ifCLASSOPTIONcaptionsoff
  \newpage
\fi

\balance
\bibliographystyle{IEEEtran}
\bibliography{main}

\end{document}